\documentclass[a4paper,11pt]{article}
\pdfoutput=1 

\usepackage{jheppub} 

\usepackage{bm,amsmath,amssymb,slashed,graphicx,%
            enumerate,alltt,xspace,multirow,xcolor,mathrsfs}
\usepackage{fancyvrb}
\usepackage{cancel}

\usepackage[normalem]{ulem}

\usepackage{subcaption}

\RequirePackage{braket}
\RequirePackage{graphicx}

\RequirePackage[numbers,sort&compress]{natbib}

\makeatletter
\g@addto@macro\bfseries{\boldmath}
\makeatother

\definecolor{labelkey}{rgb}{0,0.5,0.0}

\usepackage{listings}
\definecolor{darkgreen}{rgb}{0,0.4,0}

\newcommand\mathd{\mathrm{d}}

\newcommand{\as}{\alpha_s}
\newcommand{\aem}{\alpha}
\newcommand{\MSbar}{\ensuremath{\overline{\text{MS}}}}
\newcommand{\MSB}{\MSbar}

\newcommand\xbj{x_{\scriptscriptstyle\rm bj}}
\newcommand\mpr{m_{\rm p}}
\newcommand\ml{m_{\rm \ell}}

\newcommand{\tmop}[1]{\ensuremath{\operatorname{#1}}}
\newcommand{\muF}{\mu_{\scriptscriptstyle \rm F}}
\newcommand{\muM}{\mu_{\scriptscriptstyle \rm M}}
\newcommand{\muP}{\mu_{\scriptscriptstyle \rm pdf}}
\newcommand{\Zp}{Z^\prime}
\newcommand{\pt}{p_{\rm \scriptscriptstyle T}}

\newcommand\Plgamma{P_{\ell\gamma}}
\newcommand\Pll{P_{\ell\ell}}
\newcommand\Plgammatwo{P^{(2)}_{\ell\gamma}}
\newcommand\Pgammaq{P_{\gamma q}}

\newcommand{\kb}{{\overline{k}}}
\newcommand{\kl}{\slashed{k}}
\newcommand{\kbl}{\slashed{\kb}}
\newcommand{\rl}{\slashed{r}}
\newcommand{\ql}{\slashed{q}}
\newcommand{\ecmsq}{E^2_{\mathrm{cm}}}
\newcommand{\ecm}{E_{\mathrm{cm}}}
\newcommand{\Qsq}{Q^2}
\newcommand{\qr}{q\cdot r}
\newcommand{\pr}{p\cdot r}

\newcommand{\pq}{p\cdot q}
\newcommand{\xl}{x_{\mathrm \ell}}
\newcommand{\fl}{f_{\mathrm \ell}}
\newcommand{\fonel}{f^{(1)}_{\mathrm \ell}}
\newcommand{\zl}{z_{\mathrm \ell}}
\newcommand{\cth}{\cos \theta}

\newcommand{\LUXlep}{{\tt LUXlep}}
\newcommand{\NNPDF}{{\tt NNPDF}}
\newcommand{\Nrep}{N_{\rm rep}}
\newcommand{\hoppet}{{\tt Hoppet}}

\title{Leptons in the Proton}
\preprint{
  \begin{flushright}
    ZU-TH 13/20\\
    MPP-2020-70
  \end{flushright}
}

\author[a,b]{Luca Buonocore,}
\author[c]{Paolo Nason,}
\author[b]{Francesco Tramontano,}
\author[d]{Giulia Zanderighi}

\emailAdd{luca.buonocore@na.infn.it}
\emailAdd{paolo.nason@mib.infn.it}
\emailAdd{francesco.tramontano@cern.ch}
\emailAdd{zanderi@mpp.mpg.de}

\affiliation[a]{University of Zurich, Winterthurerstrasse  190, 8057 Zurich, Switzerland}
\affiliation[b]{Universit\`a di Napoli and INFN, Sezione di
  Napoli, Complesso Universitario di Monte Sant'Angelo,
  Via Cinthia 21, 80126 Napoli, Italy} 
\affiliation[c]{Universit\`a di Milano-Bicocca and INFN, Sezione di
  Milano-Bicocca, Piazza della Scienza 3,20126 Milano, Italy}
\affiliation[d]{Max-Planck-Institut f\"ur Physik, F\"ohringer Ring 6,
  80805 M\"unchen, Germany}

\date{Received: date / Accepted: \today}

\abstract{As is the case for all light coloured Standard Model
  particles, also photons and charged leptons appear as constituents
  in ultrarelativistic hadron beams, and admit a parton density
  function (PDF). It has been shown recently that the photon PDF can
  be given in terms of the structure functions and form factors for
  electron-proton scattering. The same holds for lepton PDFs. In the
  present work we set up a calculation of the lepton PDFs at
  next-to-leading order, using the same data input needed in the
  photon case.  A precise knowledge of the lepton densities allows us
  to study lepton-initiated processes even at a hadron collider, with
  all possible combinations of same-charge, opposite-charge,
  same-flavour, different-flavour leptons and leptons-quarks, most of
  which cannot be realized in any other forseable experiment. The
  lepton densities in the proton are extremely small, so that their
  contribution to Standard Model processes is generally shadowed by
  processes initiated by coloured partons. We will show, however, that
  there are cases where these processes can be relevant, giving rise
  to rare Standard Model signatures and to new production channels,
  that can enlarge the discovery potential of New Physics at the LHC
  and future high energy colliders with hadrons in the initial state.
}


\begin{document}


\maketitle


\newcommand{\citere}[1]{Ref.\,\cite{#1}}
\newcommand{\citeres}[1]{Refs.\,\cite{#1}}

\section{Introduction}
\label{sec:intro}

The current LHC research program is on the one hand aiming at high
precision measurements, to spot any deviations from the Standard Model,
and on the other hand at the direct search of particles arising in New
Physics scenarios.  The vast majority of New Physics searches carried
out at the LHC regards processes initiated by coloured partons, and
lepton initiated processes are relegated to future colliders involving
leptons.  On the other hand, the current LHC and its planned
high-luminosity (and eventually high-energy) upgrade is the collider
that will provide the largest part of new high-energy particle physics
data in the next 20 years, and its reach in the lepton-initiated
channels should also be exploited.  In fact, it is well known that
quantum fluctuations can give rise to the presence of leptons inside a
proton, although with a much smaller relative abundance with respect
to the coloured partons. When leptons initiate a large momentum
transfer scattering, the process becomes perturbative, the lepton
densities also obey an evolution equation, and a partonic calculation
of the process, including higher-order corrections, becomes
possible. It is thus natural to explore what is the reach of the LHC
as far as lepton-initiated processes are concerned, also considering
the fact that at the LHC the three charged leptons contribute
democratically, and thus processes (such as, for example, $\mu\tau$
scattering) that are not available at lepton colliders may be
accessed.

It is the aim of the present work to derive a precise determination of
the lepton densities inside a proton, which can then be used to
compute processes involving leptons in the initial state at hadron
colliders.

In \citeres{Manohar:2016nzj,Manohar:2017eqh} (we will refer to these
references as LUX1 and LUX2 respectively, and as ``LUX papers'' for
both of them) it was first pointed out that the parton distribution
function (PDF) of the photon in the proton can be computed with high
precision using only information from electron-proton scattering
data. In these works it was also pointed out that, for similar
reasons, this is also the case for lepton PDFs. In the present paper
we undertake the task of computing the lepton PDFs in a framework that
is very similar to the one adopted in the LUX papers, and in fact by
also heavily using the numerical code developed there.

\begin{figure}[htb]
  \begin{subfigure}{.5\textwidth}
    \centering
    \includegraphics[width=0.6\linewidth]
    {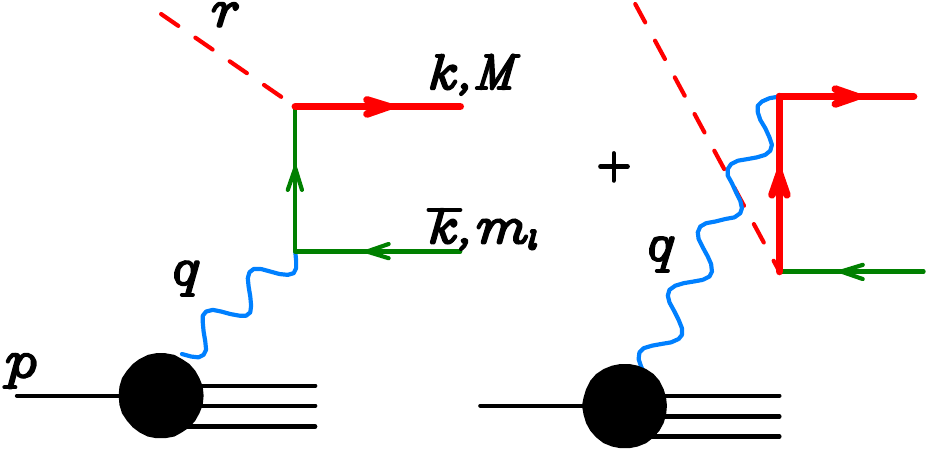}
    \caption{Structure function computation}
  \end{subfigure}
  \begin{subfigure}{.5\textwidth}
    \includegraphics[width=1\linewidth]{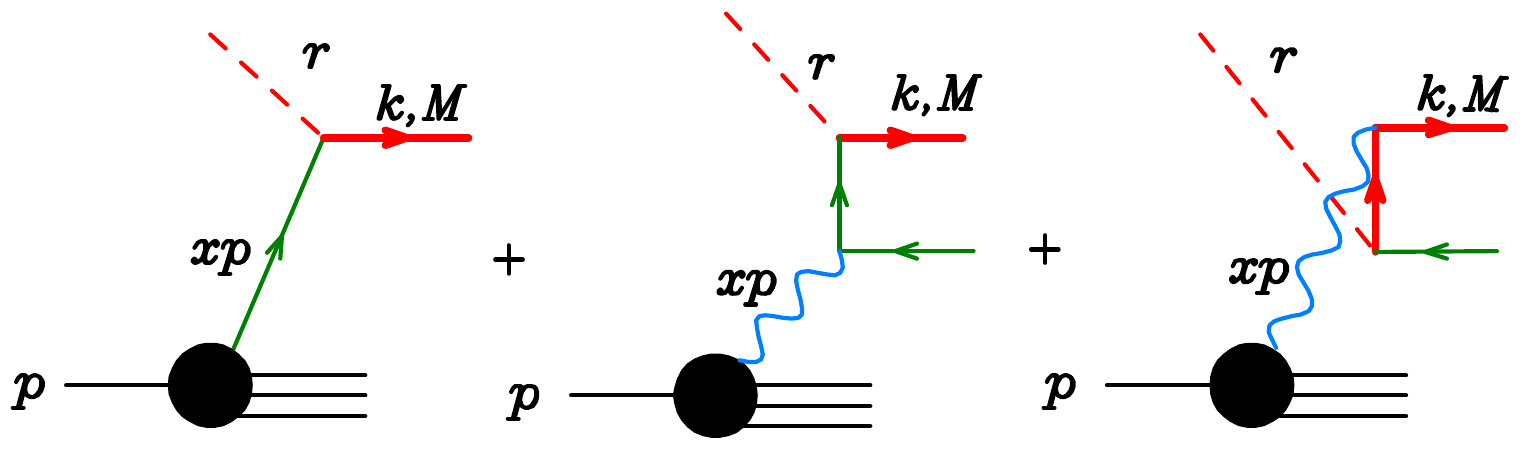}
    \caption{Parton level computation}
  \end{subfigure}
  \caption{\label{fig:basicProcess} Our basic fictitious process, with
    a scalar of momentum $r$ scattering off a light lepton and turning
    it into a heavy lepton of mass $M$, represented by the thick red
    fermion line.  In (a) we show the sum of the two diagrams that
    relate this process to the $e p$ scattering structure
    functions. In (b) we show the diagrams that enter the calculation
    of the same process at next-to-leading order according to the QCD
    factorization formula. Notice that the second and third diagram in
    (b) (unlike the (a) diagrams) are computed for an on-shell photon,
    and the collinear singularity from the photon splitting into
    leptons that arises there is subtracted.}
\end{figure}

In order to compute the lepton PDF, we consider here a fictitious
Deep-Inelastic-Scattering process involving a lepton in the proton,
i.e. the collision of a (fictitious) massless scalar with a
proton. The massless scalar interacts only with leptons, via the vertex
\begin{equation}\label{eq:bsminteraction}
  \bar{\psi}_{\mathrm{h}} \psi \phi+\mbox{c.c.},
\end{equation}
that turns the light lepton $\psi$ into a fictitious heavy lepton
$\psi_{\rm h}$, carrying a mass $M$ much larger than typical hadronic
scales. According to the parton model approach, such process can be
computed in terms of the light-lepton parton density $\fl(x)$, using
the standard QCD factorization formula. It can also be computed in
terms of the electromagnetic current structure functions. In
Fig.~\ref{fig:basicProcess}
we show schematically the representations of both computations. By
relating the two results, we can obtain the lepton parton density,
entering the second computation, in terms of the leptoproduction
structure functions, entering the first one. The factorization theorem
guarantees that the PDF so obtained is independent of the particular
process used in the calculation, as was shown explicitly in the photon
case in LUX2.

In \citere{Fornal:2018znf} the LUX method was applied to the
computation of the PDFs of the $W$ and $Z$ bosons. We stress however,
that in the case of the $W$ and $Z$, thanks to their large masses, the
electroproduction structure functions are needed only in the
perturbative regime, and thus the whole calculation can be carried out
in perturbation theory, in a way that closely resembles the
computation of the heavy flavour parton
density~\cite{Collins:1986mp}.  The case of light leptons
is instead more similar to the photon one, where the structure
functions are needed also in the very low $Q^2$ region, and thus must
be extracted from low $Q^2$ experimental data.

The paper is organized as follows. In Sec.~\ref{sec:calculation} we
present our calculation of the lepton PDFs.  We first define our
target accuracy, that is based on a careful counting scheme of the
strong and electromagnetic coupling constants and of the relevant
logarithms. We then proceed to the calculation of the lepton PDF in
the limit of zero lepton mass, first in terms of the electroproduction
structure function, and then according to the parton model formula at
next-to-leading order (NLO). We use the two results to extract a
formula for the lepton PDF. We then illustrate how the result changes
when the lepton mass is included in the calculation. Finally, we
explicitly verify that our lepton PDF satisfies the Altarelli-Parisi
evolution equation~\cite{Altarelli:1977zs}, including QED splitting
processes that also involve a term of second order in QED.

In Sec.~\ref{sec:ThErr} we explain our procedure to assess the
theoretical uncertainty of our final result, which closely follows the
one used in LUX2.  In Sec.~\ref{sec:pdfset} we describe how one can
add our lepton PDFs to any full LHAPDF set and we do this in the case
of the {\tt NNPDF31\_nlo\_as\_0118\_luxqed} set of
Ref.~\cite{Bertone:2017bme}. In Sec.~\ref{sec:validation} we show a
number of results that validate our procedure.  In
Sec.~\ref{sec:pheno} we present a number of phenomenological
applications of our lepton PDFs. In particular we consider rare SM
signatures of different flavour isolated di-lepton production; the
production of leptophilic $Z'$; the production of doubly charged
Higgs; and the production of leptoquarks. For this last case, we show
that we can reach unexplored regions of the parameter space using
already existing data from the LHC. Finally, we give our conclusions
in Sec.~\ref{sec:conclu}. In the Appendices
\ref{app:partoniccalc}-\ref{app:aem3} we provide further technical
details.

\section{Details of the calculation}
\label{sec:calculation}
We now illustrate our calculation. We first compute our probe process
in terms of the electroproduction structure functions. Then we compute
the same process in the parton model, and combine the two results to
extract the lepton PDFs.  We finally verify that our PDF satisfy the
Altarelli-Parisi equations to the appropriate order.

Before we begin, it is useful to clarify what accuracy we expect from
our calculation.  To this end, we consider the parton densities for
quarks and gluons as being of order one.  In fact, perturbatively they
may be considered as sum of terms of order $(\as L)^n$, where
$L=\log(\muF^2/\Lambda^2)$, $\muF$ is the factorization scale, and
$\Lambda$ is a typical hadronic scale.  The strong coupling $\as$ is
evaluated at a scale of order $\muF$. Thus, since $\as\approx 1/L$,
all these terms are of order 1.  NLO corrections to the quark and
gluon PDFs have the form $\as(\as L)^n$.

Besides these rules, we should also worry about the possible impact of
higher-order electromagnetic effects. These may be relevant if some
logarithmic enhancement compensates for the smallness of the
electromagnetic coupling constant.  We will adopt the same criterium
used in the LUX papers, i.e.  that $\aem$ is of the same order as
$\as^2$.  The photon density is of order $\aem L (\as L)^n$, so, in
our counting scheme, it is equivalent to $\aem L$. NLO corrections to
it are of order $\aem$ and according to our $\aem\approx \as^2$ rule,
we should also include terms of order $\aem^2 L^2$.  The lepton PDFs
are of order $\aem^2 L^2$, and their NLO corrections of order $\aem^2
L$ and $\aem^3 L^3$.  However, we will see in the following that the
terms of order $\aem^3 L^3$ are in practice much smaller than the
$\aem^2 L$ terms, and thus the $\aem\approx \as^2$ rule is a quite
conservative assumption.

In the present work we aim at NLO accuracy. The computation of the
probe process in terms of structure functions involving the graphs of
Fig.~\ref{fig:basicProcess}~(a) includes all terms of order
$\aem^2$. All possible strong corrections are already included in the
electroproduction structure functions. However, in our calculation we
also need to include NLO corrections of relative order $\aem
L$. Corrections of this kind are already present in the
electroproduction structure functions (they arise from collinear
photon radiation from quarks) and in the self-energy corrections to
the photon propagator. We can account for the latter by using the same
effective QED coupling used in the LUX papers.  The only term of
relative order $\aem L$ that we miss arises from collinear photon
radiation from the light lepton. These contributions, however, are
easily included using the evolution equations, with a method that will
be described in due time.  In the parton model calculation (the
diagrams in Fig.~\ref{fig:basicProcess}~(b)) the counting of the order
goes as follows. The first diagram is of order $\aem^2L^2$ (i.e. the
leading order of the lepton PDF). The two remaining diagrams are of
order $\aem L$ (i.e. the leading order of the photon PDF) times
$\aem$, leading to an NLO contribution of order $\aem^2L$. As one can
easily convince oneself, no other NLO corrections arise here, since in
the QCD-improved parton model calculations no large logarithms can
arise from radiative corrections.

We clarify from the start that throughout this paper we refer to the
lepton density as the density of either charge, i.e. not the sum of
lepton and antilepton, and our LHAPDF implementation returns the
density of each signed lepton. In our approximation the lepton and
antilepton densities are equal, and remain equal at higher QCD orders.
Differences arise only as subleading electroweak effects that are not
considered here.

\subsection{Calculation in terms of Structure Functions}
\label{sec:lprobe}

We begin by considering the scattering process
\begin{equation}
\phi(r) + \gamma(-q) \to \overline{\psi}_{\mathrm h}(k,M) + \psi(\kb, 0)\,, 
\end{equation}  
where $\phi(r)$ denotes a scalar of momentum $r$, $\gamma(-q)$ a
photon of momentum $-q$ ($Q^2 = -q^2$), $\overline{\psi}_{\mathrm
  h}(k,M)$ is the (hypothetical) heavy anti-lepton of mass $M$ and
momentum $k$, and $\psi(\kb, 0)$ is a massless lepton of momentum
$\kb$.  We define the kinematics of the process in terms of the
following variables
\begin{equation}
  \begin{split}
    S & = (p+r)^2\approx 2 p\cdot r\,,\\
    \ecmsq & = (r-q)^2 = -2 \qr - Q^2 \,, 
  \end{split}
\end{equation}
where $p$ is the proton momentum. 
We introduce the following dimensionless variables:
\begin{eqnarray}
  \xl &=& \frac{M^2}{S}\,,\qquad 
  x   = \frac{\ecm^2}{S}\,,\qquad 
  \zl = \frac{M^2}{\ecm^2}=\frac{\xl}{x}\,,\\
  \xbj &=& \frac{\Qsq}{2\pq}\,,\qquad 
  z   = \frac{x}{\xbj}\,.
\end{eqnarray}
In the parton model language $\xl$ can be identified with the fraction
of momentum of the lepton with respect to the proton; $x$ with the
fraction of momentum of the photon with respect to the proton; $\zl$
with the fraction of momentum of the lepton with respect to the photon
that has created it; $\xbj$ with the fraction of momentum of the quark
with respect to the proton; and $z$ with the fraction of momentum of
the photon with respect to the quark that has emitted it.

Summing the two diagrams in Fig.~\ref{fig:basicProcess} (left) we
obtain the amplitude for this process
\begin{equation}
  {\cal A}^{\mu}(r,q,k) = \bar u(k,M) \frac{ i (\kl-\rl)}{(r-k)^2}(-i e \gamma^{\mu}) v(\kb,0)  +
  \bar u(k,M) (-i e \gamma^{\mu}) \frac{ i (\kbl+\ql)}{(\kb+q)^2} v(\kb,0)  \,,
\label{eq:amp}
\end{equation}
from which, upon integration over the two-body phase space, we obtain the leptonic tensor 
\begin{equation}
  L^{\mu\nu}(r,q) = \int [\mathd\Phi_2] {\cal A}^{\mu}(r,q,k) {\cal A}^{\nu*}(r,q,k)\,,
\end{equation} 
where we have implicitly assumed the sum and averages over the spin of
the external particles.  The cross section can then be written
as\footnote{This is as Eq.~(1) in Ref.~\cite{Manohar:2016nzj}, except
  for the delta function present there that represents the
  one-particle phase space.  Here the phase space is included in $L$.}
\begin{equation}
  \sigma= \frac{1}{4\pr} \int \frac{\mathd^4q}{(2\pi)^4} \frac{1}{Q^4} L^{\mu\nu}(r,q) (4\pi) W_{\mu\nu}(p,q)\,, 
\label{eq:sigma}
\end{equation}
where $W_{\mu\nu}$ is the standard hadronic tensor, which, for the
scattering of a photon of momentum $q$ off a proton of momentum $p$, has the form 
\begin{align}
W_{\mu\nu}(p,q) &= F_1 \left(-g_{\mu\nu} + {q_\mu
q_\nu\over q^2}\right) + {F_2 \over p \cdot q} \left(p_\mu - {p\cdot q \ q_\mu\over
q^2}\right)
\left(p_\nu - {p\cdot q\ q_\nu\over q^2}\right)\,. 
\label{eq:Wmunu}
\end{align}
For ease of notation, here and in the following we will omit the
arguments of the structure functions, which will be always evaluated
at $(\xbj,Q^2)$.
We also introduce the longitudinal structure function
\begin{align}
  F_L &\equiv \left(1+\frac{4\xbj^2\mpr^2}{Q^2}\right)F_2  - 2\xbj F_1, 
\label{eq:FL}
\end{align}
which is of order ${\cal O}(\alpha_s)$ relative to $F_2$. We will
write our results using $F_2$ and $F_L$ instead of $F_2$ and $F_1$. We
remind the reader that the term of ${\cal O}(m_p^2/Q^2)$ should be kept, 
since, even if it is of higher twist in the structure functions, it
leads to leading twist contributions in the photon and lepton PDFs.

In order to make contact with the results of the LUX papers, using the identity
\begin{equation}
\int \frac{\mathd \ecmsq}{2 \pi} (2 \pi) \delta((r-q)^2- \ecmsq)=1\,,
\end{equation}  
we rewrite Eq.~(\ref{eq:sigma}) as
\begin{equation}
  \sigma= \int \frac{\mathd\ecmsq}{2\pi} \frac{1}{4\pr} \int
  \frac{\mathd^4q}{(2\pi)^4} \frac{1}{Q^4} L^{\mu\nu}(r,q) (4\pi)
  W_{\mu\nu}(p,q) (2 \pi) \delta((r-q)^2- \ecmsq)\,,
\label{eq:sigma2}
\end{equation}
where now the inner integral matches exactly Eq.~(1) of
Ref.~\cite{Manohar:2016nzj}, provided $M^2$ is replaced everywhere by
the invariant mass of the heavy-light leptons system, $\ecmsq$. Hence
we can use the same phase space as in Eq.~(3.6) of LUX2 and obtain
\begin{equation}
  \sigma= \int \frac{\mathd\ecmsq}{2\pi} \frac{1}{4\pr} \frac{1}{16\pi^2 \ecmsq}
  \int_x^{1-\frac{2x \mpr}{\ecm}} \mathd z \int_{\frac{\mpr^2
      x^2}{1-z}}^{\frac{\ecmsq(1-z)}{z}} \frac{\mathd Q^2}{Q^2}
  L^{\mu\nu}(r,q) (4\pi) W_{\mu\nu}(p,q) \,.
\label{eq:sigma3}
\end{equation}
Note that the variable $x$ defined in the LUX papers as $x=M^2/S$
should be replaced here by $x=\ecmsq/S$.

For the leptonic tensor, writing explicitly the $\mathd \Phi_2$ phase
space, we obtain
\begin{equation}
  L^{\mu\nu}(r,q) = \frac{1}{16\pi} \left(1-\frac{M^2}{\ecmsq}\right)
  \int \mathd \cth {\cal A}^{\mu} {\cal A}^{\nu *}\,,
\end{equation} 
where $\theta$ is the angle between $k$ and $r$ in the centre-of-mass (CM) of the
scalar-photon system.
The leptonic tensor is gauge invariant, and hence it can be written as
\begin{align}
  L_{\mu\nu}(r,q) &= L_1 \left(-g_{\mu\nu} + {q_\mu q_\nu\over q^2}\right)
  + {L_2 \over r \cdot q} \left(r_\mu - {r\cdot q \ q_\mu\over
q^2}\right)
\left(r_\nu - {r\cdot q\ q_\nu\over q^2}\right)\,, 
\label{eq:Lmunu}
\end{align}
where $L_1, L_2$ are functions of $\zl$, $Q^2$ and $M^2$. 

At this point, we have all the elements to compute $L^{\mu\nu} W_{\mu
  \nu}$ in terms of the proton structure functions. The expression
obtained is rather lengthy, hence we do not report it here.  We note
however that the result simplifies considerably if we only retain the
terms that are relevant to our approximation.  It turns out that the
expression for $L^{\mu\nu} W_{\mu \nu}$ has schematically the form
\begin{equation} \label{eq:LWschematic}
  L^{\mu\nu} W_{\mu\nu}=  F \times P(S,M^2,Q^2,\ecmsq,\mpr^2) L(\ecmsq,M^2,Q^2)+ F \times R(S,M^2,Q^2,\ecmsq,\mpr^2),
\end{equation}
where $P$ and $R$ are rational functions of their arguments.  We have
indicated schematically with $F$ the linear dependence of the result
upon the structure functions. Furthermore we have defined
\begin{eqnarray}\label{eq:Logdef}
  L(\ecmsq,M^2,Q^2)&=&\log\frac{M^2}{Q^2}+\log\frac{(\ecmsq-M^2)\ecm^4}{M^6} \nonumber \\
  &+&   \log\frac{(\ecm^4+Q^2(\ecm^2-M^2)) (\ecm^2-M^2+Q^2)}{\ecm^4(\ecm^2-M^2)}\,.
\end{eqnarray}
The $\log (M^2/Q^2)$ arises in the leptonic tensor from the integral in $\mathd \cos\theta$.
In fact, it is easy to see that the first diagram in
Fig.~\ref{fig:basicProcess}~(a), in the limit of small $Q^2$
has a collinear divergence when the anti-lepton is produced in the forward
direction.

The $P$ and $R$ coefficients can be separated in the following terms:
\begin{enumerate}
\item Terms that behave as $\mpr^2/Q^2$ for small $Q^2$.
\item Terms that do not depend upon $Q^2$.
\item Terms that vanish at small $Q^2$.
\end{enumerate}
The terms of the first item give rise to an integral of the form
\begin{equation}
  \int \frac{\mathd Q^2}{Q^4} \mpr^2
\end{equation}
multiplying the structure functions, and, in the case of $P$, by
$L$. Neglecting the mild $Q^2$ dependence in the structure functions
and in $L$, this integral is dominated by small values of $Q^2 \approx
\mpr^2$ (since the lower limit of integration in Eq.~(\ref{eq:sigma3})
is proportional to $\mpr^2$).  Thus, the contribution proportional to
$P$ is of order $\log M^2/\mpr^2$ (arising from the $L$ coefficient),
while the one proportional to $R$ is of order 1.  According to our counting,
we would only need to keep
the former, that in our approximation is NLO. However,
we will also keep the latter, that is of order NNLO.
The reason for doing this will be clarified in due time.
 
 The terms of the second item give clearly origin to single and double
 logarithmic enhancement in the contributions proportional to $P$, and
 to single logarithm enhancement in the contributions proportional to
 $R$.

 The terms of the third item lead to integrals (dominated by large
 values of $Q$) that are of order one, and thus negligible in our
 approximation.  We also notice that the third logarithm in
 Eq.~(\ref{eq:Logdef}) vanishes for small $Q^2$, and thus can be
 neglected for the same reason.  With these simplifications the cross
 section in Eq.~\eqref{eq:sigma3} becomes
\begin{equation}
  \begin{split}
\sigma  &= \frac{\pi}{M^2} \left(\frac{\aem}{2\pi}\right)^2 \int_{\frac{M^2}{S}}^1 d
\zl \int_x^1 \frac{dz}{z} \int_{\frac{\mpr^2
    x^2}{1-z}}^{\frac{\ecmsq(1-z)}{z}} \frac{dQ^2}{Q^2} \\ & \Bigg\{
\Plgamma(\zl) \left[ F_2 \left(z \Pgammaq(z) + \frac{2 \mpr^2
    x^2}{Q^2}\right) - F_L z^2\right]\log\frac{M^2(1-\zl)}{\zl^3 Q^2}
 \\ & + F_2 \left[
    4(z-2)^2\zl(1-\zl)- z \Pgammaq(z) \right]
 +F_Lz^2 \Plgamma(\zl)-\frac{2\mpr^2 x^2}{Q^2}F_2
  \Bigg\} \,,
  \end{split}
\label{eq:sigmafin}
\end{equation}
where we have introduced the $q\to q\gamma$ and $\gamma \to \bar{l} l$
splitting functions
\begin{equation}
  \Pgammaq(z) = \frac{1+(1-z)^2}{z} \,, \qquad 
  \Plgamma(\zl) = 1-2\zl+2\zl^2 \,.
\end{equation}  
We stress again that in the above expression the structure functions
are evaluated at $(\xbj, Q^2)$.  We notice that in the last line of Eq.~(\ref{eq:sigmafin})
there is a term proportional to $F_L$ that is not multiplied by a
large logarithm. Since $F_L$ is down by one power of $\as$ with
respect to $F_2$, this term leads to a contribution that is subleading
in our counting scheme. However, we keep it, together with the last term,
proportional to $F_2$, that is dominated by small value of $Q^2$. Thus, the only terms
that we have dropped in our calculations are those that are dominated by large value of $Q^2$,
and yield subleading contributions proportional to a  structure function evaluated at
a large scale multiplied by $\aem^2$.

\subsection{The parton model calculation}

We now present the result for the computation of the same cross
section using a parton model calculation. The equivalence between the
two expressions will allow us to derive a formula for the lepton PDF
in terms of the hadronic structure functions. The details of the
partonic calculation are reported in
Appendix~\ref{app:partoniccalc}. The final result is
\begin{eqnarray}
  \frac{\sigma}{\sigma_B} & = & \int \mathd x \fl (x, \mu_F^2) \delta (S x -
  M^2) + \frac{\aem}{2 \pi}  \frac{1}{M^2} \int^1_{\frac{M^2}{S}} \mathd x f_{\gamma} (x, \mu_F^2) 
 \nonumber\\   & \times &\left\{ \zl \Plgamma(\zl) \left[ \log \frac{M^2}{\mu_F^2} + \log \frac{(1 -
  \zl)^2}{\zl^2} \right] + 4 \zl^2 (1 - \zl) \right\}, 
\label{eq:sigpartonic} 
\end{eqnarray}
where $\zl$ is now given as a function of $x$, $\zl = M^2 /
E_{\tmop{cm}}^2 = M^2 / (S x)$, and $\sigma_B = \pi$.

\subsection{Extraction of the lepton PDF }
\label{sec:lpdf}

In order to extract the lepton PDF we identify the two expressions for
$\sigma$ in Eq.~\eqref{eq:sigmafin} and Eq.~\eqref{eq:sigpartonic}.
We obtain
\begin{eqnarray}
  \xl{}\fl (\xl{}, \mu_F^2)&=& 
                           M^2 \int_0^1 \mathd x \fl (x, \mu_F^2) \delta (S x - M^2) \nonumber \\
  & = & - \frac{\aem
  (\mu_F^2)}{2 \pi}  \int^1_{\xl}  \mathd x f_{\gamma} (x)  \left\{ \zl{} \Plgamma
  (\zl{}) \left[ \log \frac{M^2}{\mu_F^2} + \log \frac{(1 - \zl{})^2}{\zl^2} \right]
  + 4 \zl^2 (1 - \zl{}) \right\} \nonumber\\
  & + & \left( \frac{1}{2 \pi} \right)^2 \int^1_{\xl}  \frac{\mathd
  x}{x} \zl{} \int_x^1 \frac{\mathd z}{z} \int_{\frac{\mpr^2 x^2}{1 -
  z}}^{\frac{E_{\tmop{cm}}^2 (1 - z)}{z}} \frac{\mathd^{} Q^2}{Q^2} \aem^2
  (Q^2) \nonumber\\ \times
  &  \Bigg\{ & \Plgamma (\zl{})  \left[ F_2  \left( z \Pgammaq (z) +
  \frac{2 \mpr^2 x^2}{Q^2} \right) - F_L z^2 \right] \log \frac{M^2 (1 -
  \zl{})}{\zl^3 Q^2}  \nonumber\\
  & + & F_2 \left[ 4 (z - 2)^2 \zl{} (1 - \zl{}) - z \Pgammaq(z)
         \right]   +F_Lz^2 \Plgamma(\zl)-\frac{2\mpr^2 x^2}{Q^2}F_2 \Bigg\}\,,  \label{eq:fl}
\end{eqnarray}
where we have replaced $\mathd\zl = \zl dx/x $, and used $\xl=M^2/S$.
We recall that $F_2, F_L$
are evaluated at $(\xbj, Q^2)$, with $\xbj = x/z$. 
We now recall the expression for the photon PDF, Eq.(6) of Ref.~\cite{Manohar:2016nzj}: 
\begin{eqnarray}\label{eq:gammalux}
  x f_{\gamma} (x) &=& \frac{1}{2 \pi \aem(\mu_F^2)} \int_x^1 \frac{\mathd
   z}{z} \Bigg\{ \int_{\frac{x^2 \mpr^2}{1 - z}}^{\frac{\mu_F^2}{1 - z}}
   \frac{\mathd^{} Q^2}{Q^2} \aem^2 (Q^2) \left[ \left( z P_{\gamma q} (z) +
       \frac{2 x^2 \mpr^2}{Q^2} \right) F_2 - z^2 F_L \right]\nonumber \\
&& \phantom{aaaaaaaaaaaaaaaaaaaaaaaaa}  - \aem^2 (\mu_F^2)   z^2 F_2(x/z, \muF^2) \Bigg\}\,, 
\end{eqnarray} 
where the structure functions in the square bracket are evaluated at $(x/z, Q^2)$,
while in the last term of the curly bracket $F_2$ is evaluated at  $(x/z, \muF^2)$.

We notice that the last term (called ``the \MSB{} correction'' in the LUX papers),
when introduced in the parton model cross section formula,
has precisely the form of the terms that have been neglected in eq.~(\ref{eq:sigmafin}),
i.e. it is of order $\aem^2$ and multiplies a structure function evaluated at a large scale.
We thus neglect it.
We also observe that if we replace the upper limit in the $Q^2$
integration in Eq.~(\ref{eq:fl}) with $\mu_F^2 / (1 - z)$ we obtain an
equivalent expression up to subleading terms, since the difference
only involves values of $Q^2$ near the upper limit, and thus again has the form
of structure functions evaluated at a large scale times $\aem^2$ (the explicit
logarithm is of order one for these values of $Q^2$).
Proceeding in this way, and substituting
Eq.~(\ref{eq:gammalux}) for $f_\gamma$, we obtain
\begin{eqnarray}
&& \xl{} \fl (\xl{}, \mu_F^2)  =  \left( \frac{1}{2 \pi} \right)^2 \int_{\xl{}}^1
  \frac{\mathd x}{x} \zl{} \int_x^1 \frac{\mathd z}{z} \int_{\frac{\mpr^2 x^2}{1
  - z}}^{\frac{\mu_F^2}{1 - z}} \frac{\mathd^{} Q^2}{Q^2} \aem^2 (Q^2) \nonumber \\
&& \phantom{aaaaa} \times  \Bigg\{  \Plgamma (\zl{}) \log \frac{\mu_F^2}{(1 - \zl{}) \zl{}
  Q^2}   \left\{ F_2  \left( z \Pgammaq (z) +
  \frac{2 \mpr^2 x^2}{Q^2} \right) - F_L z^2 \right\}  \nonumber \\
&& \phantom{aaaaaa} +   F_2 \left[ 4 (z - 2)^2 \zl{} (1 -
   \zl) - (1+4\zl(1-\zl)) z \Pgammaq (z) \right]
   \nonumber \\ 
  && \phantom{aaaa}+F_Lz^2 \Plgamma(\zl)-\frac{2\mpr^2 x^2}{Q^2}F_2
- \left(F_2 \frac{2\mpr^2x^2}{Q^2} - z^2 F_L\right)  4  \zl  (1-\zl)
  \Bigg\} .  \label{eq:leptonpdf0}
\end{eqnarray}
where we have used $\xl{} = M^2 / S$ and $\zl{} = \xl{} / x$. The
structure functions $F_2$ and $F_L$ are always evaluated at $\xbj=x/z$
and $Q^2$.

The above formula has been derived in the massless limit for the
physical lepton.  As shown in Appendix~\ref{app:masseffects}, the
effect of lepton masses are simply
accounted for by replacing
\begin{equation}\label{eq:mleffect}
\log \frac{\mu_F^2}{(1 - \zl{}) \zl{} Q^2} \rightarrow \log\frac{\muF^2}{(1-\zl)\zl \left(Q^2+\frac{\ml^2}{\zl(1-\zl)}\right)}\,,
\end{equation}
and adding a term propotional to the square of the lepton mass.
We thus obtain
\begin{eqnarray}
  \xl{} \fl (\xl{}, \mu_F^2) & = & \left( \frac{1}{2 \pi} \right)^2 \int_{\xl{}}^1
  \frac{\mathd x}{x} \zl{} \int_x^1 \frac{\mathd z}{z} \int_{\frac{\mpr^2 x^2}{1
                               - z}}^{\frac{\mu_F^2}{1 - z}} \frac{\mathd^{} Q^2}{Q^2} \aem^2 (Q^2)
  \nonumber \\
                       & \Bigg\{&\Plgamma (\zl{})
        \log\frac{\muF^2}{(1-\zl)\zl \left(Q^2+\frac{\ml^2}{\zl(1-\zl)}\right)}
  \left[ F_2  \left( z \Pgammaq (z) +
   \frac{2 \mpr^2 x^2}{Q^2} \right) - F_L z^2 \right]
  \nonumber \\  \label{eq:leptonpdf1}
  & + &  F_2 \left[ 4 (z - 2)^2 \zl{} (1 -
        \zl{}) - \left(1+4\zl{}(1-\zl{})\right) z \Pgammaq (z) \right] \nonumber \\
 \nonumber \\ 
 &+& F_Lz^2 \Plgamma(\zl)-\frac{2\mpr^2 x^2}{Q^2}F_2 -
                                 \left(F_2 \frac{2\mpr^2x^2}{Q^2} - z^2 F_L\right)  4  \zl  (1-\zl) \nonumber \\ 
&+& \frac{ \ml^2F_2}{\ml^2+Q^2\zl(1-\zl)} 
   \left[ z\Pgammaq(z) -8\zl(1-\zl)\left(1-z-\frac{\mpr^2x^2}{Q^2}\right)
   +\frac{2\mpr^2 x^2}{Q^2}    \right] \nonumber \\
&-& \frac{\ml^2F_Lz^2}{\ml^2+Q^2\zl(1-\zl)}\left[2- \Plgamma(\zl)\right]        \Bigg\}\, .
\end{eqnarray}
This is our final expression for the lepton PDF. It can be evaluated
numerically, similarly to what was done recently for the photon
PDF. We observe that, compared to the latter, the lepton PDF requires
one extra integration.

The structure functions $F_2$ and $F_L$ are the only functions in
formula~(\ref{eq:leptonpdf1}) that are not known analytically. They
depend only upon the two variables $\xbj$ and $Q^2$. It is thus
possible to express formula~(\ref{eq:leptonpdf1}) as an integral in
$\xbj$ and $Q^2$, and a third variable, from which the integrand
depends analytically.  In Appendix~\ref{app:twoDimIntegrand} we
provide some details regarding this simplification of the integrand.
The integration in the third variable is thus simpler to perform,
either with numerical methods (e.g.  using Gaussian integration) or
analytically.
\subsubsection{Subleading $\aem^2$ terms}
In our calculation of the lepton PDF we have kept some subleading terms, that are formally of order $\aem^2$ (and thus contribute at the NNLO level) 
but are dominated by values of $Q^2$ that are much lower than the high scale of the process (i.e. $M^2$, $\muF^2$ or $\ecm^2$,
that in our calculation should be considered of the same order). The motivation for doing so is better illustrated
if we examine what we would need to extend the accuracy of our calculation to include terms of order $\aem^2$.
First of all, in the partonic formula, the NLO term should be evaluated with an $f_{\gamma}$ that is accurate at the
NLO level. The only NLO term that we dropped from $f_{\gamma}$ is the \MSB{} subtraction term, that leads to a correction
of order $\aem^2 F_2$, with $F_2$ evaluated at the scale $\muF^2$.
Next, in the calculation in terms of structure functions we have dropped
all non-singular terms at small $Q^2$. These terms are dominated by
values of $Q^2$ of the order of the high scales and thus have the form
of a structure function at large scale times $\aem^2$. Furthermore, we
have dropped terms having to do with the upper limit of the $Q^2$
integration that also have the form $\aem^2 F_{2/L}$. Finally, we
should add the higher-order graphs to the parton model formula. The
only missing graph that can contribute at NNLO is the collision of the
massless scalar with a quark, producing a heavy lepton and the light
(physical) antilepton. This graph is of order $\aem^2$, and it
multiplies a quark parton density evaluated at a scale $Q^2$. Thus,
\emph{the only contributions of higher order that we have omitted have
  the form $\aem^2$ multiplied by some parton densities evaluated at a
  high scale}. Under these circumstances, it makes sense to estimate
the uncertainty due to missing higher order terms by scale variation
methods. These in fact only involve changes in the region of $Q^2\approx
\muF^2$. The subleading terms that we have included, on the other
hand, all involve values of $Q^2\ll \muF^2$. Thus, they have little
sensitivity to the value of $\muF$. One can easily check, by taking a
logarithmic derivative of our lepton PDF with respect to $\muF^2$,
that their $\muF$ dependence is either power suppressed (as is the
case for the terms suppressed by $\mpr$ or by the lepton mass) or is
of order $\aem^2 \as$ (as is the case for the $F_L$ terms).  Thus,
they do not influence the scale dependence of the result, and, if we
don't include them, scale variation cannot be used to assess their
order of magnitude, and we would have to estimate the error associated
with their absence in some different way. On the other hand, we can
easily include all of them, and do not need to worry about the error
due to their absence.

As a further observation to confirm the soundness of this procedure, we notice that as we go to higher orders, no
more terms of this kind arise in either the calculation in terms of structure functions, or in the partonic calculation. This being the case, these
terms should be process independent, since parton density functions are process independent.  This is apparent if we look at the terms proportional to powers of the lepton masses. They are characterized
by configurations with small $Q^2$ for the photon, and small virtuality of the lepton arising from the photon splitting.
Configurations of this kind factorize in terms of the Born scalar-lepton cross section, that we divide out to obtain
the lepton PDF. Thus, their contribution does not depend upon the process. For the remaining terms, they are still
dominated by small $Q^2$, but the lepton arising from photon splitting may have both large and small virtuality. On the other hand,
terms of this kind arise in the partonic cross section carried out up to the NLO level, and their combination with the
direct calculation should yield a result that is again process independent, and only dependent upon the adopted
subtraction procedure.

\subsection{Verifying the Altarelli-Parisi evolution}\label{sec:AP}
The lepton PDF given in formula~(\ref{eq:leptonpdf1}) must satisfy the
Altarelli-Parisi equation~\cite{Altarelli:1977zs}. Given that
Eq.~(\ref{eq:leptonpdf1}) includes accurately terms of order $\aem^2
L^2$ and $\aem^2 L$, where $L=\log(\muF/\Lambda)$, its logarithmic
derivative must contain accurately terms of order $\aem^2 L$ and
$\aem^2$.  Taking the derivative of formula~(\ref{eq:leptonpdf1}) we
obtain
\begin{eqnarray}
  \frac{\partial \xl{} \fl (\xl{}, \mu_F^2)}{\partial \log \mu_F^2} & = & \left(
  \frac{1}{2 \pi} \right)^2 \int_{\xl{}}^1 \frac{\mathd x}{x} \zl{} \int_x^1
  \frac{\mathd z}{z} \int_{\frac{\mpr^2 x^2}{1 - z}}^{\frac{\mu_F^2}{1 - z}}
   \frac{\mathd^{} Q^2}{Q^2} \aem^2 (Q^2)
   \nonumber \\ & \times &
  \Plgamma (\zl{})  \left\{ F_2  \left( z \Pgammaq (z) +
  \frac{2 \mpr^2 x^2}{Q^2} \right) - F_L z^2 \right\}  \nonumber \\
   & + & \left( \frac{\aem \left(\mu_F^2 \right)}{2 \pi} \right)^2 \int_{\xl{}}^1
  \frac{\mathd x}{x} 
  \zl{} \int_x^1 \frac{\mathd z}{z}
  \Bigg\{ \left[ \Plgamma \left(\zl{} \right) \log
  \frac{(1 - z)}{(1 - \zl{}) \zl{}} \right] F_2\, z \Pgammaq (z)  \nonumber \\
  & + & F_2 [4 (z - 2)^2 \zl{} (1 - \zl{}) - (1 + 4 \zl{} (1 - \zl{})) z \Pgammaq
  (z)] \Bigg\}, \label{eq:Ap0}
\end{eqnarray}
where in the first term we have taken a derivative with respect to the
explicit $\mu_F^2$ dependence of the logarithm, while in the second term we have
taken a derivative with respect to the upper limit of the integration. In doing so
we have neglected terms of order $\ml^2/\muF^2$ that arise in the argument
of the logarithm, and we have replaced
\begin{equation}
  \aem^2 \left(\frac{\mu_F^2}{1-z} \right) =  \aem^2 \left(\mu_F^2 \right) +
  {\cal O}(\aem^3),
\end{equation}
and neglected the higher-order terms.  In the first term of
Eq.~(\ref{eq:Ap0}), the $z$ integral corresponds to the LUX expression
of the photon parton density (Eq.~(\ref{eq:gammalux})), except that it
does not include the term outside the $Q^2$ integral (this was
referred to as the $\overline{\tmop{MS}}$ correction in the LUX
papers). We can thus replace this expression with the photon parton
density, adding a term to compensate for the lack of the
$\overline{\tmop{MS}}$ correction. We get
\begin{eqnarray*}
  \frac{\partial \xl{} \fl (\xl{}, \mu_F^2)}{\partial \log \mu_F^2} & = & \left(
  \frac{\aem (\mu_F^2)}{2 \pi} \right) \int_{\xl{}}^1 \frac{\mathd x}{x} 
  \zl{} \Plgamma (\zl{}) \left[ x f_{\gamma} (x) + \frac{\aem (\mu_F)}{2\pi}
  \int_x^1 \mathd z z F_2 \right]\\
   & + & \left( \frac{ \aem \left(\mu_F^2 \right)}{2 \pi} \right)^2
      \int_{\xl{}}^1 \frac{\mathd x}{x} 
        \zl{}\, \int_x^1 \frac{\mathd z}{z}
        \Bigg\{ \left[ \Plgamma \left(\zl{}\right) \log
  \frac{(1 - z)}{(1 - \zl{}) \zl{}} \right] z \Pgammaq (z)  \\
  & + & [4 (z - 2)^2 \zl{} (1 - \zl{}) - (1 + 4 \zl{} (1 - \zl{})) z \Pgammaq
  (z)] \Bigg\} F_2\\
  & = & \left( \frac{\aem (\mu_F^2)}{2 \pi} \right) \xl{}
  \int_{\xl{}}^1 \frac{\mathd x}{x} \Plgamma (\zl{}) f_{\gamma} (x)\\
  & + & \left( \frac{ \aem \left(\mu_F^2\right)}{2 \pi} \right)^2 \xl{}
  \int_{\xl{}}^1 \frac{\mathd x}{x^2} 
   \int_x^1 \frac{\mathd z}{z} \Bigg\{ \left[
  \Plgamma \left(\zl{} \right) \log \frac{(1 - z)}{(1 - \zl{})
  \zl{}} \right] z \Pgammaq (z)  \\
  & + & \Plgamma (\zl{}) z^2 + [4 (z - 2)^2 \zl{} (1 - \zl{}) - (1 + 4 \zl{} (1 -
  \zl{})) z \Pgammaq (z)]  \Bigg\} F_2,
\end{eqnarray*}
where in the second equality we have transferred the subtracted
$\overline{\tmop{MS}}$ correction to the second term.  We now rewrite
the integral of the second term as
\begin{eqnarray}
 \int_{\xl{}}^1 \frac{\mathd x}{x^2} \int_x^1 \frac{\mathd z}{z}
   &=& \int_0^1 \mathd \xbj \int_{\xl{}}^1 \frac{\mathd x}{x^2} 
   \int_x^1 \frac{\mathd z}{z} \delta \left( \xbj -
  \frac{x}{z} \right)
  \nonumber \\
  &=& \int_{\xl{}}^1 \frac{\mathd
   \xbj}{\xbj} \int_{\xl{}}^1 \frac{\mathd x}{x^2} 
    = \int_{\xl{}}^1 \mathd \xbj
  \int_{\frac{\xl{}}{\xbj}}^1  \frac{\mathd z}{x^2},
\end{eqnarray}
where now $x = z \xbj$, and replace $F_2$ with
\begin{equation}
  F_2 = \xbj \sum c_q^2 f_q .
\end{equation}
Defining $\xi = \xl{} / \xbj$, we find
\begin{eqnarray}
&&  \frac{\partial \fl (\xl{}, \mu_F^2)}{\partial \log \mu_F^2}  = 
  \frac{\aem (\mu_F^2)}{2 \pi}  \int_{\xl{}}^1 \frac{\mathd x}{x}
  \Plgamma (\zl{}) f_{\gamma} \left(x,\muF^2\right) + \left( \frac{\aem (\mu_F)}{2 \pi}
  \right)^2 \int_{\xl{}}^1 \frac{\mathd \xbj}{\xbj}
   \Bigg\{ - (1 + \xi) \log^2 \xi \nonumber \\
  &+& \frac{(8 \xi^2 + 15 \xi +
  3)}{3} \log \xi + \frac{(1 - \xi) (28 \xi^2 + \xi + 10)}{9 \xi} \Bigg\}
  \sum c_q^2 f_q\left(\xbj,\muF^2\right)\,.
  \label{eq:plq_splitting}
\end{eqnarray}
The expression in the curly bracket is equal to the function $p_s
(\xi)$ in Eq.(60) of Ref.~\cite{deFlorian:2016gvk}, that enters the
$P_{lq}$ splitting function. We stress that, in our case, this term is
of order $1/L$ relative to the first term. In fact, $f_\gamma$ is of
order $\aem L$, so that the first term is of order $\aem^2 L$, while
the second term is of order $\aem^2$, since $f_q$ is of order 1 in our
counting scheme. Of course there are other terms of order $\aem^2$ in
the second order QED evolution, but they multiply either $f_\gamma$ or
$\fl$, and are thus subleading in our counting scheme.  We also remind
the reader that our expression for the lepton PDF does not include
terms of order $\aem^3 L^3$, that, if present, would give rise to the
term
\begin{equation}
  \frac{\aem(\muF^2)}{2\pi}\int_{\xl{}}^1 \frac{\mathd x}{x} P_{ll}(\zl{})\fl\left(x,\muF^2\right)
\end{equation}
on the right-hand side of Eq.~(\ref{eq:plq_splitting}).  This term is
of order $\aem^3L^2$ in our counting scheme ($\fl$ is of order $\aem^2
L^2$).  Since we assume $L^{-2}$ to be of order $\aem$, this amounts
to a NLO correction to the evolution that should be present.

\section{Theoretical error due to missing higher-order effects}\label{sec:ThErr}
In order to estimate the theoretical error due to missing higher-order
effects, we parallel the method proposed in LUX2. There, the upper
limit in the $Q^2$ integration yielding the LUX photon parton density
was modified using a generic $z$-dependent form $M^2(z)$ (see sec. 9.1
in LUX2).  This modification was compensated by a corresponding
modification of the \MSbar{} conversion term. As we will discuss in
the following section, our determination of the lepton densities is
performed together with a determination of the photon density. We thus
apply the same method to our lepton-density formula
Eq.~(\ref{eq:leptonpdf1}), replacing the upper integration limit
$\muF^2/(1-z)$ with $M^2(z)$. In the lepton case, this variation is
already of one order above our target accuracy, and does not give rise
to any modification of the \MSbar{} conversion term. Thus, following
LUX2, we consider two forms for $M^2(z)$
\begin{equation}
  M^2(z)=\frac{\muM^2}{1-z}\quad \mbox{and} \quad  M^2(z)=\muF^2,
\end{equation}
and take $\muM$ to be a multiple of $\muF$, to be varied by a factor
of two above and below $\muF$. The corresponding range of results is
our estimate for the theoretical error due to missing higher-order
effects.

\section{Construction of a PDF set with leptons}
\label{sec:pdfset}
We now illustrate our construction of a PDF set including photons and
leptons.  This set is based upon the {\tt
  NNPDF31\_nlo\_as\_0118\_luxqed} set of Ref.~\cite{Bertone:2017bme},
and will be made available as an LHAPDF set under the name {\tt
  LUXlep-NNPDF31\_nlo\_as\_0118\_luxqed}. Here we refer to it simply
as the \LUXlep{} set. For brevity, in the following, we will also
refer to the NNPDF set upon which it is based as the \NNPDF{} set.

The construction of a PDF set with leptons relies upon
electroproduction data, both for the elastic and the inelastic case.
The same data and fits used in the LUX papers (see
Refs.~\cite{Bernauer:2013tpr,Lee:2015jqa,
  Airapetian:2011nu,Osipenko:2003bu,Christy:2007ve,Abramowicz:1991xz,Abe:1998ym,Ricco:1998yr,
  Liang:2004tj}) are used here.

We constructed a PDF set with leptons starting from the \NNPDF{} set.
For better consistency, we generate
the photon ourselves using the LUX approach.  We implemented our
formula for the leptons, Eq.~(\ref{eq:leptonpdf1}), by suitably extending the computer code
developed for the LUX papers.

We proceeded as follows.
\begin{enumerate}
\item\label{enum:start} We compute the lepton and photon densities at
  a reference scale $\muF$ (our central value will be $\muF=20\,{\rm
    GeV}$), using formula~\ref{eq:leptonpdf1}, as a function of the
  electromagnetic structure functions and form factors. The structure
  functions in the perturbative regime are evaluated using a member
  $m$ of the \NNPDF{} set.
\item We take the member $m$ at a reference scale $\muP$ and evolve it
  to $\muF$ using \hoppet~\cite{Salam:2008qg}.  The photon evolution
  is included at order $\aem$ and $\aem\as$, but leptons are not
  included in any splitting function. This evolution step matches what
  is done in the \NNPDF{} set, where leptons where simply not
  included. We do this step ourselves for better stability of the
  results.

  We did not take $\muP$ equal to the initial NNPDF evolution scale
  $\muF^{(0)}=1.65\,$GeV in order to avoid an excessive sensitivity to
  the evolution implementation in NNPDF. In fact, we want to use
  \hoppet{} for the evolution, and eventual subleading differences
  with the evolution implemented in NNPDF would manifest themselves
  especially at low scales, where lower and higher-orders effects
  become closer in size. On the other hand, we cannot take $\muP$
  arbitrarily large, since the backward evolution of the Altarelli-Parisi
  equations is unstable, and generates unphysical oscillations.
  We have found that the choice  $\muP=7\,$GeV is a good compromise,
  since it allows to evolve the set backward down to a value
  of $\muF^{(0)}=2\,$GeV without visible oscillations,
  and since the value of the coloured parton
  densities at larger scales are still consistent with those
  of \NNPDF{}.\footnote{We stress that this is a technical problem
  related to our use of an evolution code that is different from
  the one used in the original PDF set.}
\item In the NNPDF set evaluated at the scale $\muF$, we replace the
  photon density with the one we computed, and add our computed
  leptons densities.
\item\label{enum:backevol} Using \hoppet, we evolve the set so
  obtained down to the initial scale $\muF^{(0)}$. This step of
  evolution includes the QED splitting functions at order $\aem$ and
  $\aem\as$, excluding those involving a lepton radiating a photon. We
  do this because in our calculation of the lepton PDFs, photon
  radiation from leptons (that is of order $\aem^3 L^3$ in our
  counting scheme) is not included, while, in our approximation, it
  should be (see the discussion in the introduction of
  section~\ref{sec:calculation}). We implement this radiation by
  evolving the lepton density from the scale $\muF^{(0)}$ using the
  full QED evolution. In order to do this we need the lepton densities
  at the low scale, and we obtain them with the procedure we just
  outlined. Notice that we cannot compute the lepton densities
  directly at a low scale, because, at low scales, our calculation is
  not guaranteed to satisfy the Altarelli-Parisi equation due to the
  power suppressed effects it includes. We thus perform the
  computation at a scale that is large enough for power suppressed
  effects to be negligible, and use the QED evolution (without lepton
  radiation, since this is not included in our calculation) to produce
  a partonic lepton density (free from power suppressed effects) at
  the low scale.

  In addition, we also added to \hoppet{} the splitting function
  $P_{lq}$, which is of order $\aem^2$, and included its effects at
  this step. Notice that the inclusion of the $P_{lq}$ splitting
  function is mandatory to preserve our accuracy. In fact, as already
  stated in sec.~\ref{sec:AP}, $P_{lq}$ contributes at the NLO level
  to the evolution of the lepton densities, since it multiplies
  directly a quark density, that it is of order one in our counting
  scheme. The leading term in the evolution is of order $\aem$ times a
  photon density (that is of order $\aem/\as$) and is thus of order
  $\aem^2/\as$. Thus the $P_{lq}$ contribution, of order $\aem^2$, is
  down by a single power of $\as$ with respect to the dominant term.
\item \label{enum:end} Starting with the set at the low scale obtained
  in this way, we generate the PDFs at any scales using \hoppet{} with
  full $\aem$ and $\aem\as$ plus $P_{lq}$ evolution including leptons.
\end{enumerate}

The procedure of item~\ref{enum:backevol} can be avoided by evaluating
directly the $\aem^3 L^3$ contribution arising from lepton
electromagnetic radiation. This calculation is illustrated (and
compared with the method of item~\ref{enum:backevol}) in
Appendix~\ref{app:aem3}. In the appendix it is also shown that this
effect is small, relative to the NLO effects (of order $\aem^2 L$)
that are included in our formula~(\ref{eq:leptonpdf1}). Thus, the rule
$\aem\approx \as^2$ that we have adopted in our counting seems to be
quite conservative, and omitting the $\aem^3 L^3$ would only lead to a
minor error.

The procedure illustrated in items \ref{enum:start} to \ref{enum:end}
is applied to all members of the PDF set. Furthermore we also apply it
to the central ($m=0$) set modified with the addition of the
uncertainty variations described in LUX2 in section 10.2, labelled as
(EFIT), (EUN), (RES), (R), (M), (PDF), (T) and (HO). We briefly
summarize their meaning.
\begin{itemize}
\item[(EFIT)] The uncertainty on the elastic contribution induced
  by the fit of the form factors \cite{Bernauer:2013tpr,Lee:2015jqa},
  as was done in~LUX2.

\item[(EUN)] The uncertainty that comes from replacing the fit
  to the elastic form factors~\cite{Bernauer:2013tpr}
  including polarisation data with the fit with only
  unpolarised data, as in~LUX2.

\item[(RES)] We replace the CLAS resonance-region fit~\cite{Osipenko:2003bu} with the
  Christy-Bosted fit~\cite{Christy:2007ve}, as in~LUX2.

\item[(R)] A modification of $R$ (the ratio of the longitudinal
  to the transverse electroproduction cross section) by $\pm 50\%$
  around its central value, as in~LUX2.
  
\item[(M)]  A modification of the $Q^2_\text{PDF}$ scale which governs the
  transition from fitted data for $F_2$ and $F_L$ to a PDF-based
  evaluation, as in~LUX2.

\item[(PDF)] The input PDF uncertainties for $Q^2 > Q_\text{PDF}^2$ according
  to the default prescription for NNPDF, as described in the following.

\item[(T)] A twist-4 modification of $F_L$, as in~LUX2.

\item[(HO)] An estimate of missing higher-order effects,
  as described in the following.
\end{itemize}

For the higher-order uncertainties (HO) we cannot use the method
described in LUX2, since we did not perform an NNLO calculation of the
lepton densities. We thus consider the scale variations described in
sec.~\ref{sec:ThErr} with the following five choices:
$M^2(z)=\muM^2/(1-z),$ with $\muM=2\muF,\muF/2$, plus $M^2(z)=\muM^2$
with $\muM=2\muF,\muF,\muF/2$, and then take the largest deviation
from the central set (in absolute value) as further uncertainty
variation.  We now must implement our variations in a way that is
consistent with the meaning of the NNPDF members, i.e. as replicas. We
have used the following approach. We assume that each of our
variation, suitably symmetrized, corresponds to a Gaussian error, with
variance equal to the variation itself. Thus, for the generic member
$m$ of the NNPDF set (excluding $m=0$, i.e. the central one), for
flavour $i$ and given $x$ and $\muF$ values, we compute the correction
\begin{equation}
\Delta_i^{(m)}(x,\muF)= \sum_{j=1}^7
\frac{f_{i,(j)}^{(0)}(x,\muF)-f_i^{(0)}(x,\muF)}{f_i^{(0)}(x,\muF)}
f_i^{(m)}(x,\muF) \times R(m,j),
\end{equation}
where $ f_i^{(m)}$ stands for the density of parton $i$ in the member
$m$, and $ f_{i,(j)}^{(m)}$ stands for the same member evaluated
according to the $j^{\rm th}$ variation among the seven possibilities
(EFIT), (EUN), (RES), (R), (M), (T) and (HO) mentioned earlier.
$R(m,j)$ is a Gaussian random number (depending only upon the set and
the variation kind) with zero average value and unit variance. We then
redefine
\begin{equation}\label{eq:newreplicas}
  f_i^{(m)}(x,\muF) \rightarrow f_i^{(m)}(x,\muF)+\Delta_i^{(m)}(x,\muF)-\frac{1}{\Nrep}\sum_{k=1}^{\Nrep} \Delta_i^{(k)}(x,\muF),
\end{equation}
where $\Nrep$ is the total number of replicas (100 in the set we are
considering). Eq.~(\ref{eq:newreplicas}) guarantees that the average
of all replicas remains the same, i.e.\ equal to the central member.

We have chosen to compute the lepton and photon PDFs at the initial
scale $Q=20\,{\rm GeV}$. We do not add an error associated with this
choice, since it is much smaller than our estimate of the error due to
higher-order perturbative effects.  In particular, choosing the much
higher value $Q=100\,{\rm GeV}$, we get a variation of the electron
density at 100~GeV below 0.5\%{} across the whole $x$ range.

\section{Validation}
\label{sec:validation}
\begin{figure}[htb]
  \centering
  \includegraphics[width=0.7\textwidth]{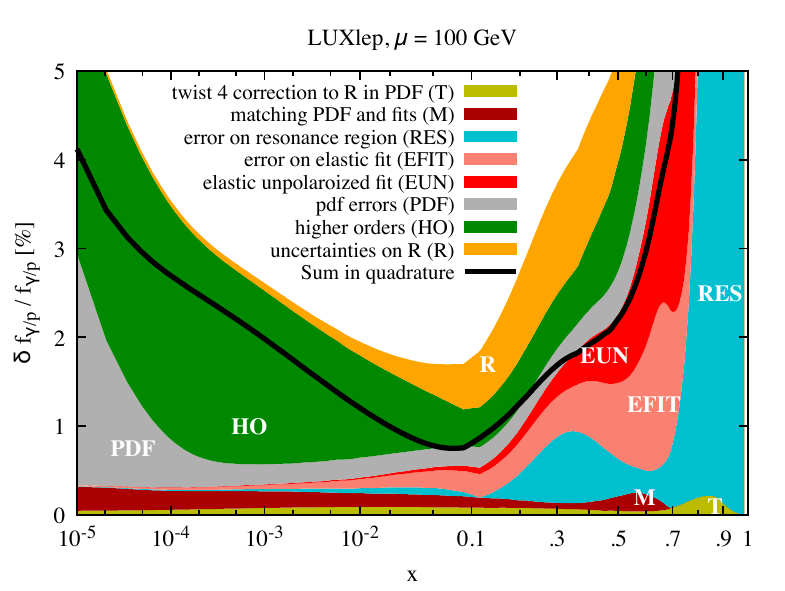}
  \caption{\label{fig:PhotonUncertainties} The relative uncertainties
    on the photon density, stacked linearly. The various entries are
    explained in the text. The black line is the sum in quadrature of
    all uncertainties.}
\end{figure}

We illustrate now how our sources of uncertainties affect the lepton
densities. First of all, however, we want to show that the photon PDF,
that we compute here using the NLO LUX approach in conjunction with
the NNPDF set, has uncertainties that are consistent with what we
found in the original LUX papers. The errors are reported in
Fig.~\ref{fig:PhotonUncertainties} at the reference scale $\mu = 100$
GeV.
The PDF uncertainty is obtained with the usual method required by PDF sets with $\Nrep$ replicas. One defines
\begin{equation}
  \Delta_i(x,\mu)=\sqrt{\frac{\sum_{j=0}^{\Nrep}(f_i^{(j)}(x,\mu)-f_i^{(0)}(x,\mu))^2}{\Nrep}}\,,
\end{equation}
by summing in quadrature the deviation of the photon density of each
replica with respect to the central set, dividing the sum by the
number of replicas, and then taking the square root.  We see that this
figure compares well with Fig.~15 of LUX2, the only marked difference
being the size of the HO uncertainty, that is much smaller there. On
the other hand, this difference is justified if we look at Fig.~13 of
the same reference, where both the uncertainty bands of the NLO and
NNLO computation of the photon density are illustrated.

\begin{figure}[htb]
  \centering
  \includegraphics[width=0.49\textwidth]{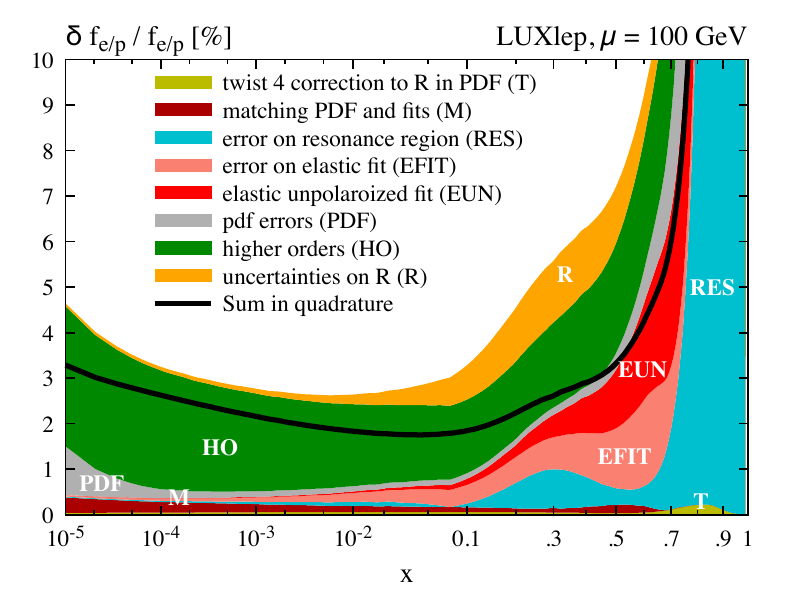}
  \includegraphics[width=0.49\textwidth]{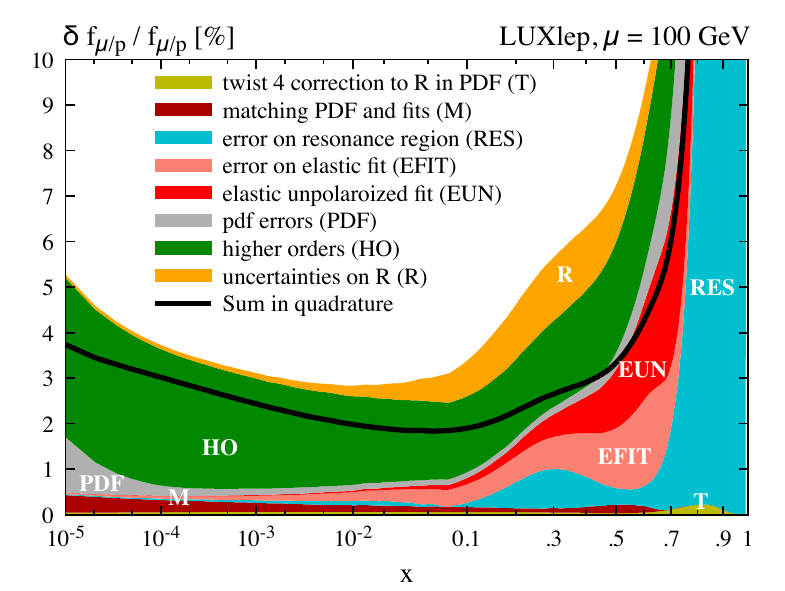}\hfill
  \includegraphics[width=0.49\textwidth]{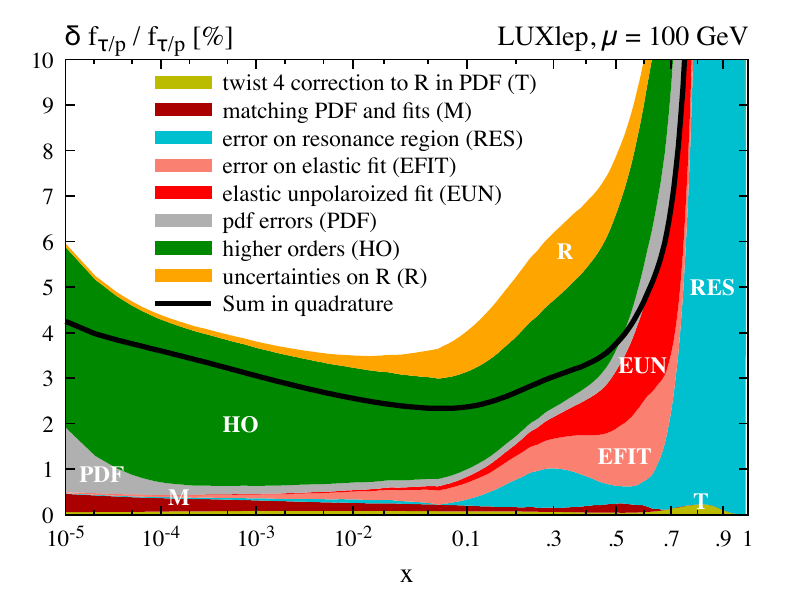}
  \caption{\label{fig:ElectronUncertainties}
    As in Fig.~\ref{fig:PhotonUncertainties} for the electron (top left), for the muon (top right) and for the tau (bottom).}
\end{figure}

In Fig.~\ref{fig:ElectronUncertainties}
we show the uncertainties for the electron, the muon and the tau at the same reference scale.
We see that the uncertainties are in line with those of the photon PDF
illustrated in Fig.~\ref{fig:PhotonUncertainties}, except for the
error due to HO effects, that seems larger for leptons, and slightly
larger for larger lepton masses.

\begin{figure}[htb]
  \centering
  \includegraphics[width=0.49\textwidth]{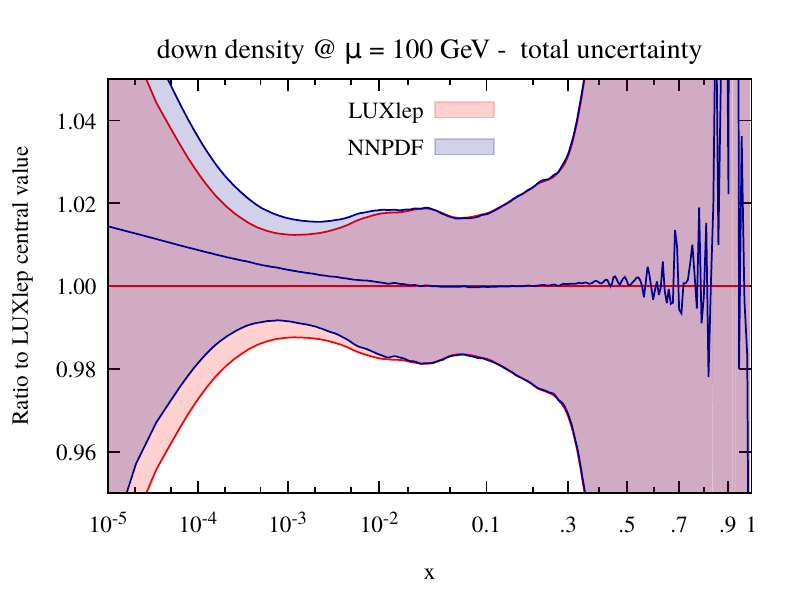}
  \includegraphics[width=0.49\textwidth]{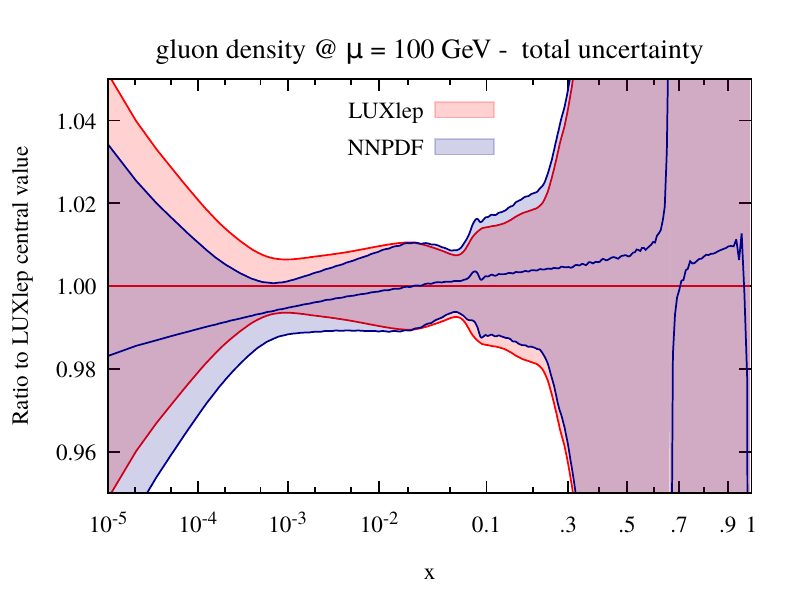}
  \caption{\label{fig:LUXlepNNPDF} The full uncertainty band of the
    quark down (left) and gluon (right) densities in the \LUXlep{} and
    in the \NNPDF{} sets, normalized to the central value of
    \LUXlep{}. The uncertainty band is obtained with the standard
    procedure that is adopted with replicas.}
\end{figure}

We now turn to the validation of the \LUXlep{} set.  In order to test
the consistency of our evolution machinery, we checked that the QCD
partons are consistent within errors in the \LUXlep{} and \NNPDF{}
sets.  In Fig.~\ref{fig:LUXlepNNPDF}
we show the comparison for the down quark and the gluon density.  The
central values and the uncertainty bands nicely overlap in the
central $x$-region. We observe the largest deviation in the
gluon case, where the central value is displaced by roughly $1\%$
at $x=10^{-3}$, with the \NNPDF{} central value touching the edge of
the \LUXlep{} band. We stress that these differences are mostly due
to our use of an evolution code that differs from the one used in the
\NNPDF{} set. The influence of the inclusion of leptons on the
distribution of coloured partons is instead
negligible, and in particular gives a
negligible contribution to the total proton momentum (we find a
momentum sum equal to $1.00065$ at $\muF=100$~GeV). We thus, at
variance with the LUX papers, did not apply any correction to restore
the momentum sum rule.

\begin{figure}[htb]
  \centering
  \includegraphics[width=0.6\textwidth]{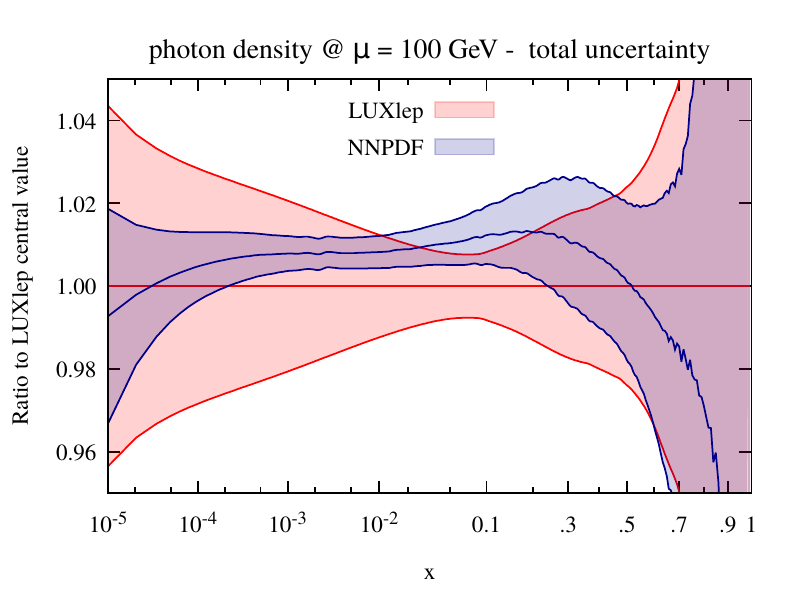}
  \caption{\label{fig:LUXlepNNPDFGamma}
    The full uncertainty band of the photon density in the \LUXlep{} and in the \NNPDF{} sets,
    normalized to the central value of \LUXlep{}. The uncertainty band is obtained with the standard
    procedure that is adopted with replicas.}
\end{figure}

In Fig.~\ref{fig:LUXlepNNPDFGamma} we compare the photon densities in
the \LUXlep{} set and in \NNPDF{} sets.
We find a reasonable agreement, with the two uncertainty bands nicely
overlapping within few percents.  The \LUXlep{} band is larger since for
consistency we recompute the photon PDF with the LUX approach at NLO
while in the NNPDF set it was obtained at NNLO. This difference
explains also the small deviation of the central values in the
central- and high-$x$ region, with NNPDF being bigger.

\begin{figure}[htb]
  \centering
  \includegraphics[width=0.49\textwidth]{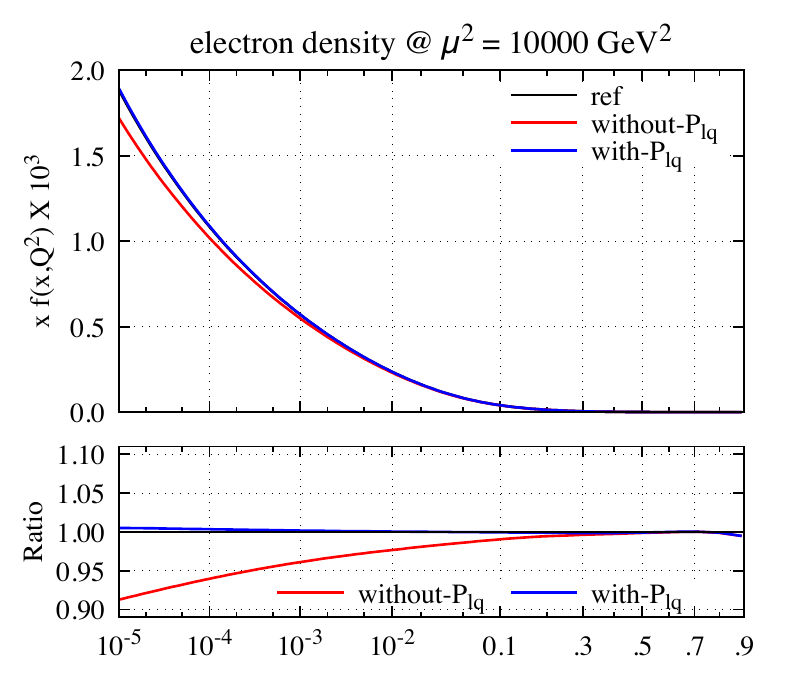}
  \includegraphics[width=0.49\textwidth]{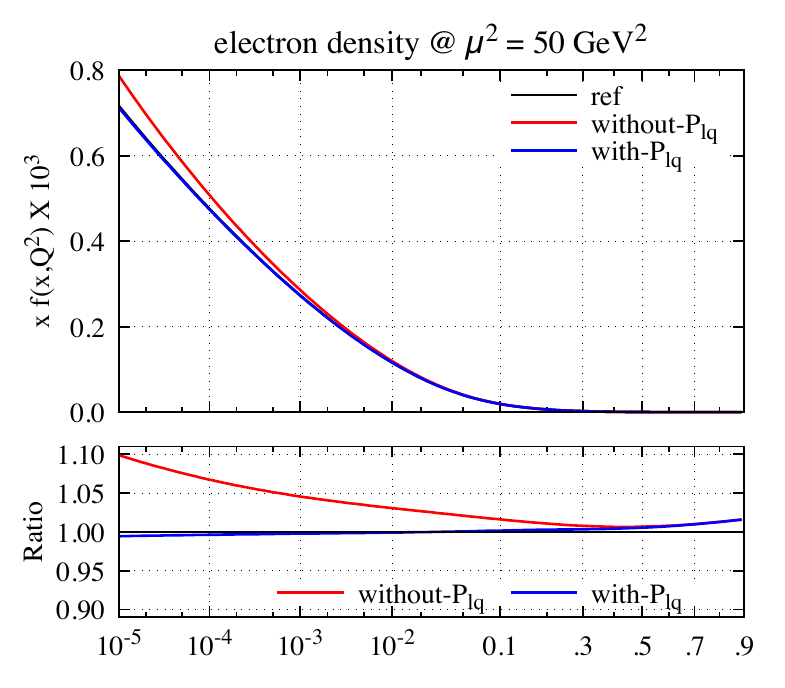}
  \caption{\label{fig:Plqimpact}
    Impact of the $P_{lq}$ splitting function in the evolution of the electron density.
    The reference curve (black line) is the lepton PDF as directly computed
    at the given scale $\mu^2$ with the LUX approach.
    The other two curves are obtained by starting with the lepton  PDF directly computed
    at the reference scale with the LUX approach, and then evolved
    either turning off (red curve) or turning on (blue curve) the $P_{lq}$ splitting.
    We show results both for the forward evolution at the scale $\mu^2=10000\,$GeV$^2$ (left panel)
    and  for the backward evolution, $\mu^2=50\,$GeV$^2$ (right panel).}
\end{figure}

We conclude this section discussing the impact of the $P_{lq}$
splitting in the evolution of the lepton densities.  We compare, for
different scale choices, the lepton densities directly computed with
the LUX approach, with the ones obtained by using the LUX approach at
a fixed reference scale (chosen to be as before $\mu^2 =
400\,$GeV$^2$), and then evolving with \hoppet{}~\cite{Salam:2008qg}
at arbitrary scales.  To be consistent with our computation of the
lepton densities, we turn off the photon emission from leptons.  We
consider two evolution options: ``without-$P_{lq}$'' which does not
include the quark-to-lepton splitting and ``with-$P_{lq}$'' which
does. In Fig.~\ref{fig:Plqimpact} we report the results for the
electron densities at the scales $\mu^2 = 10000\,$GeV$^2$ and $\mu^2 =
50\,$GeV$^2$.
These plots clearly show the relevance of the inclusion of the
$P_{lq}$ splitting, which is crucial to achieve the NLO accuracy. The
effects are especially large in the small-$x$ region where the
$P_{lq}$ splitting gets logarithmic enhanced contributions (see
Eq.~\eqref{eq:plq_splitting}) and the omission of the $P_{lq}$
splitting would lead to deviations of order $10\%$ for the scales
considered.

In Ref.~\cite{Bertone:2015lqa} a study of lepton PDFs was performed,
and a PDF set with lepton was obtained. This study was carried out
before the LUX procedure was available. Large variations were found
there depending upon the assumptions on the initial conditions for the
photon and lepton densities. Nevertheless, at large factorization
scales, these studies should capture at least the order of magnitude
of the lepton and photon densities, since they are prevalently
generated by perturbative radiation. We have compared our set with the
set of Ref.~\cite{Bertone:2015lqa}, and have found that indeed they
are compatible in order of magnitude, with differences that range from
10\%{} in the small-$x$ region, up to 50\%{} for large $x$. These
findings are in line with the fact that there is a contribution to the
large-$x$ photon PDF coming from the low $Q^2$ region (see Fig.~18 of
LUX2) that amounts to about 50\%{} of the total, and can only be
computed with reliable accuracy by exploiting the electron scattering
data as we do.

\section{Phenomenology}
\label{sec:pheno}

The precise determination of the leptonic content of the proton allows
us to consider the LHC also as either a (broad band beams) high energy
lepton-(quark/gluon) or a lepton-lepton collider, even including muons
and taus in the initial state, which are beyond the current collider
accelerator technology. In the next subsections, after a brief
illustration of the associated luminosities, we present some
physically motivated applications of lepton-initiated processes at the
LHC.

\subsection{Lepton luminosities}
\begin{figure}[t]
  \centering
  \includegraphics[width=\textwidth]{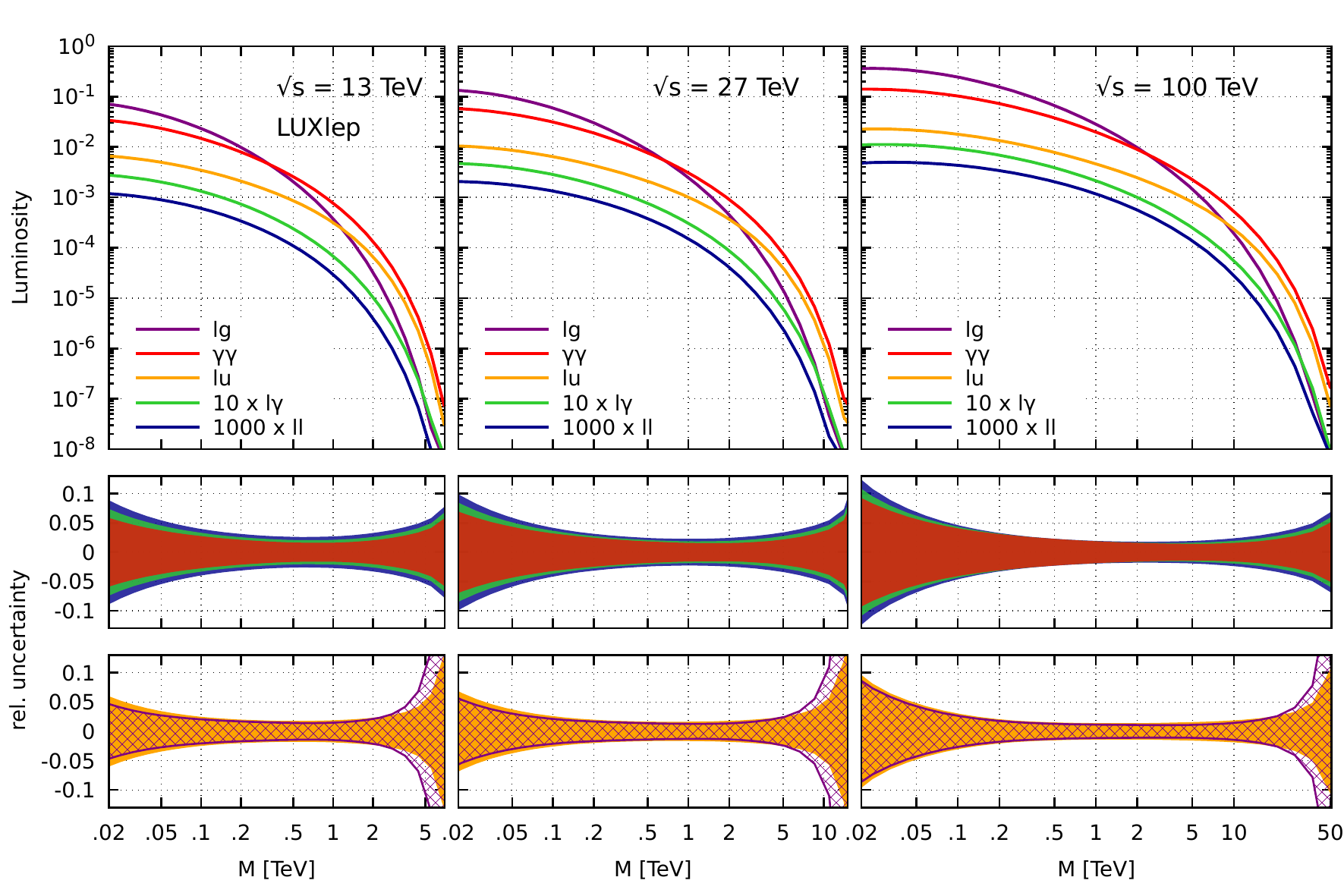}
  \caption{ The lepton-gluon ($lg$, purple), photon-photon ($\gamma
    \gamma$, red), lepton-up ($lu$, orange), lepton-photon ($l\gamma$,
    green), and lepton-lepton ($ll$, blue) luminosities at 13~TeV
    (left), 27~TeV (center) and 100~TeV (right) in $pp$ collisions (as
    defined in the text) in the case of the electron, computed using
    the \LUXlep{} set. Differences with respect to the muon or tau case
    cannot be appreciated on the scale of the plot.
 \label{fig:lumi}}
\end{figure}

We begin by showing in Fig.~\ref{fig:lumi}
the luminosities, defined as
\begin{eqnarray}
  {\cal L}_{i j} &\equiv&  M^2 \int_0^1 \mathd z\, \mathd y\,  f_i(z,M^2)
  f_j\left(y,M^2\right) \delta(M^2-s zy) \nonumber\\
  & = &
  \frac{M^2}{s}\int \frac{\mathd z}{z} f_i(z,M^2)
  f_j\left(\frac{M^2}{zs},M^2\right)\,,
\end{eqnarray}
that we computed for $pp$ collisions at 13~TeV (left), 27~TeV (middle)
and 100~TeV (right) using the \LUXlep{} set. The error bands are obtained
with the standard method used for PDFs with replicas. In particular we
show lepton-gluon ($lg$, purple, ${\cal L}_{l^-g}+{\cal L}_{g\,l^-}$),
lepton-up ($lu$, orange, ${\cal L}_{l^-u}+{\cal L}_{u\,l^-}$)
lepton-photon ($l\gamma$, green, ${\cal L}_{l^-\gamma}+{\cal
  L}_{\gamma\, l^-}$) and lepton-lepton ($ll$, blue, ${\cal
  L}_{l^+l^-}+{\cal L}_{l^-l^+}$) luminosities.
As a reference, we also show the photon-photon luminosity ($\gamma
\gamma$, red, ${\cal L}_{\gamma\gamma}$).  In the upper panels we only
show the central values, since the uncertainty band is too small to be
appreciated.  In the two bottom panels we show the relative
uncertainties, obtained with the usual prescription adopted in sets
with replicas (see Sec.~\ref{sec:pdfset} and \ref{sec:validation}).
The uncertainties are all very similar and below 5\% over a large
range of $M$. Only for very high masses ($M/\sqrt{s} \gtrsim 0.3$) the
uncertainties exceed 5\%. In the case of the $lu$ and $lg$
luminosities it is clear that the uncertainty is dominated by the
uncertainty on the QCD partons. If we compare the $ll$ and $lu$
luminosities we note that the former is suppressed by a factor of
about $~8\cdot 10^{3}$ with respect to the latter. Similarly the
$l\gamma$ luminosity is suppressed by a factor of about $200$-$300$
with respect to the $\gamma \gamma$ luminosity.

\begin{figure}[htb]
  \centering
  \includegraphics[width=\textwidth]{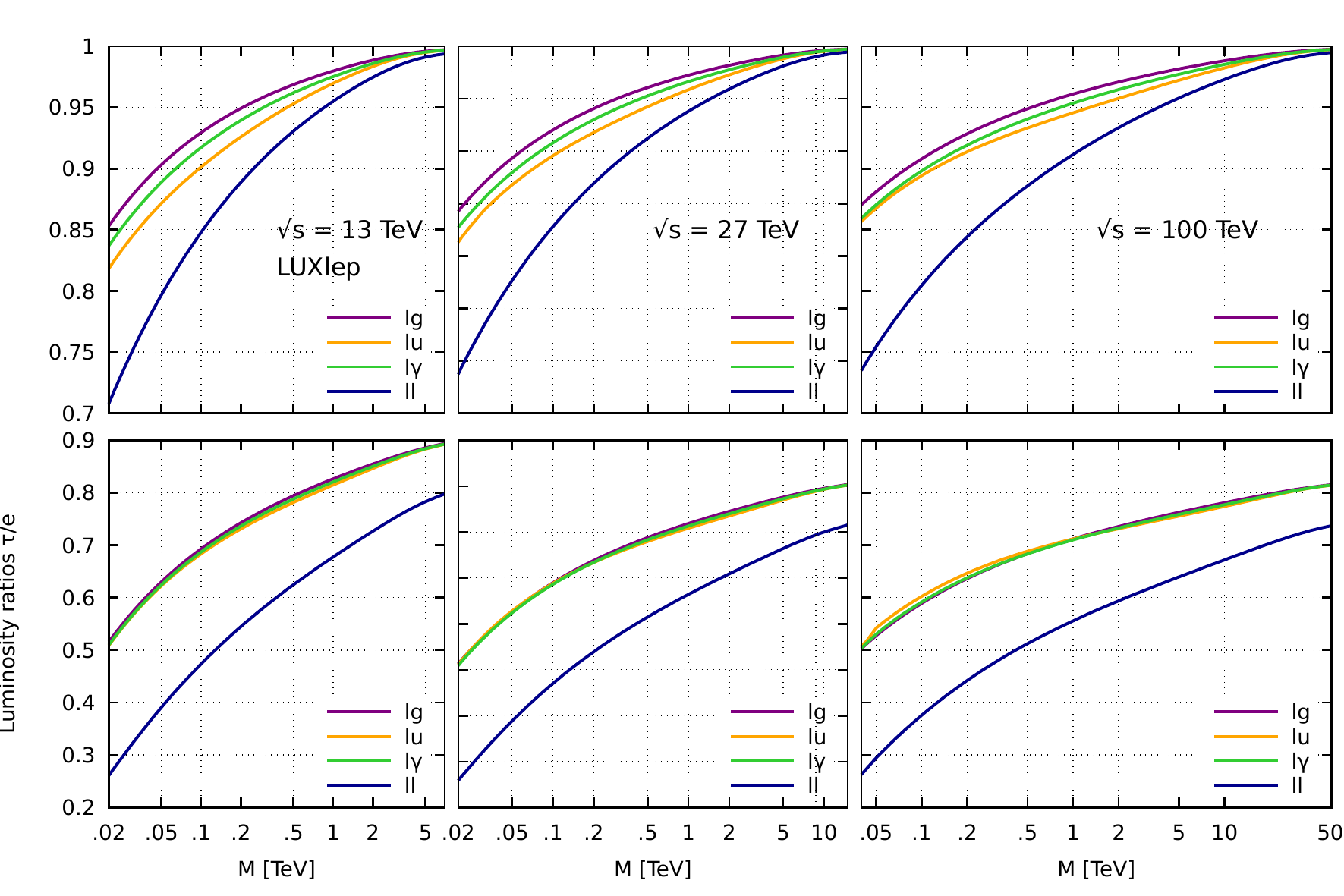}
  \caption{ The ratio of the lepton-gluon ($lg$, purple),
    photon-photon ($\gamma \gamma$, red), lepton-up ($lu$, orange),
    lepton-photon ($l\gamma$, green), and lepton-lepton ($ll$, blue)
    luminosities involving muons (upper panels) or taus (bottom
    panels) to the same luminosities involving electrons, at 13 TeV
    (left), 27 TeV (center) and 100 TeV (right).
    \label{fig:lumi-ratio}}
\end{figure}

Next we show in Fig.~\ref{fig:lumi-ratio}
the ratio of the luminosities involving taus and muons to the ones
involving electrons plotted in Fig.~\ref{fig:lumi}.  As expected, the
luminosities of heavier leptons are of the same order of magnitude as
the ones involving electrons, but they are somewhat suppressed, in
particular at lower masses.

\subsection{NLO corrections}
We remark that, in order to exploit the accuracy of our lepton PDFs,
NLO calculations of the lepton-initiated processes are needed. In
general, these will involve processes where incoming leptons are
replaced by incoming photons splitting into a lepton pair,\footnote{We
  stress that these processes are not of higher order in QED, as one
  may naively think.  Thus, we expect them to be of the same order as
  typical NLO QCD corrections.}  where the associated collinear
singularity is subtracted.  These subprocesses are suppressed by a
single power of $L$, i.e. the logarithm of the ratio of the process
scale to some typical hadronic scale. This is because the photon PDF
is larger than the lepton PDF by a factor $1/(L\aem)$, and the NLO
correction carries an extra factor of $\aem$ and no large logarithms
(since the collinear divergence has been subtracted). In the following
applications, that are given for illustrative purposes, these
higher-order corrections are not included.

We can expect that, at the LHC, lepton-initiated processes may become
competitive with other production mechanisms for the search of New
Physics objects that have a preferential coupling to leptons.  Since
the New Physics objects we are searching for are expected in general
to be very massive, one may also worry about the fact that our lepton
PDFs are computed including a photon exchange, but no $Z$ exchange
diagrams. A back of the envelope estimate of these effects would lead
to an increase of the lepton PDFs of the order of 5\%{} at TeV
scales. On the other hand, our definition of the lepton PDFs without
$Z$ contributions can consistently be used, as long as $Z$ exchange
effects are included as higher-order corrections, that, due to the $Z$
mass, do not present collinear singularities. Thus, the effect of the
inclusion of $Z$ exchange should be considered together with the
inclusion of NLO effects, that we are neglecting in the following
examples.

\subsection{Lepton-lepton scattering}
Signatures that at the LHC have been considered exotic so far, and
important to test flavour violating interactions, are two isolated,
back-to-back leptons of different flavours with the same or with
opposite charge (see
e.g. Refs.\cite{ATLAS:2014kca,Aaboud:2017qph,Sirunyan:2018xiv,Sirunyan:2020ztc}).
Since our parton densities now include lepton PDFs, we are in a
position to estimate the Standard Model (SM) contribution to these
signatures coming from $\ell \ell' \to \ell \ell'$ scattering mediated
by a photon. These SM processes are accompanied by no other
significant activity in the event.

We consider here both 13 and 27 TeV collisions and require standard
transverse momentum and rapidity cuts on the leptons,
\begin{equation}
  p_{t,\ell} > 20  {\rm GeV}\,, \qquad |\eta_{\ell}| < 2.4\,.
\label{eq:cuts}
\end{equation}
Since the processes we are considering are dominated by a photon
exchange in the $t$-channel, we set the factorization scale to the
lepton transverse momentum, and estimate the uncertainty on the cross
sections by varying the factorization scale by a factor of 2 up and
down. 
  In Tab.~\ref{tab:sigma-llp}
\begin{table}[thpb]
\begin{center}
\begin{tabular}{|c|c|c|c|c|c|c|}
\hline
\rule{0pt}{2ex}    & $e^+ \mu^-$  &$e^+ \tau^-$  &$\mu^+ \tau^-$  & $e^+e^+$& $\mu^+ \mu^+$  &$\tau^+ \tau^+$  \\
\hline 
\rule{0pt}{3ex} $\sigma_{\rm 13 TeV}$ [fb] & $0.29^{+0.13}_{-0.10}$ & $0.18^{+0.11}_{-0.08}$ & $0.16^{+0.10}_{-0.07}$ & $0.24^{+0.10}_{-0.08}$ & $0.19^{+0.09}_{-0.07}$ & $0.08^{+0.06}_{-0.04}$ \\ 
\hline
\rule{0pt}{3ex} $\sigma_{\rm 27 TeV}$ [fb] & $0.53^{+0.25}_{-0.18}$ & $0.34^{+0.21}_{-0.15}$ & $0.30^{+0.19}_{-0.14}$ & $0.440^{+0.19}_{-0.14}$ & $0.34^{+0.16}_{-0.12}$ & $0.14^{+0.12}_{-0.07}$ \\ 
\hline 
\end{tabular}
\end{center}
\caption{Standard Model production cross sections for same sign and
  opposite sign leptons of different flavours at 13 and 27
  TeV.
 The uncertainty is obtained by varying the central factorization
 scale, set to the lepton transverse momentum, by a factor of 2 up and
 down.
  \label{tab:sigma-llp}}
\end{table}
we give cross sections for the hard scattering among different lepton
flavours for both 13 and 27 TeV proton-proton collisions. Rates for
charged-conjugated processes are identical.  The large errors are due
to the fact that the results that we present here are only LO
accurate, but the calculation can be easily carried out at NLO order
in QCD.  One can see that at the end of the High Luminosity program,
with an estimated integrated luminosity of 3 ab$^{-1}$, about 850 $e^+
\mu^-$, 550 $e^+ \tau^-$ and 500 $\mu^+ \tau^-$ will be produced and
pass our basic lepton cuts. At 27 TeV the cross-sections are almost a
factor two higher, and the luminosity is a factor of five higher, so
that we expect roughly 10 times more events.  The dominant SM
background for same sign leptons comes from $W^+W^+$ (or $W^-W^-$)
plus dijet production. For example, the fully inclusive leading order
cross section for $W^+W^+$, with the $W$-bosons decaying into a single
lepton species is about 443~fb at 13~TeV. However, if one requires
that the leptons pass the cuts of Eq.~\eqref{eq:cuts}, vetoes on jets
with $p_{t,j} > 20$ GeV, requires a missing transverse momentum less
than $p_{\rm t,miss} = 15$ GeV and further requires the two leptons to
be balanced in transverse momentum and back to back
\begin{equation}
  \frac{|\vec{p}_{t,{\ell_1}} +\vec{p}_{t,{\ell_2}}|}{\max\{p_{t,{\ell_1}},p_{t,{\ell_2}}\}} < 0.1\,, \qquad 
  |\Delta \phi_{{\ell_1} {\ell_2}} - \pi | < 0.1\,,  
\end{equation}
then the background coming from $W^+W^+$ decays reduces to about 0.1
ab. This means that not a single event is expected to pass the cuts
even at the end of the HL-LHC program. These additional cuts have
instead no effect on the cross-sections quoted above.
  
Besides same sign $W$ pairs, one should also consider the very
abundant background coming from heavy flavour production ($c$ and $b$
flavoured hadrons) that decay leptonically. In order to get rid of
these backgrounds, it is crucial to estimate to what extent one can
veto events for additional hadronic activity, that is bound to be much
smaller for lepton-initiated processes. This requires the availability
of a shower Monte Carlo that can handle incoming leptons.

\subsection{$\Zp$ searches}
\begin{figure}
  \centering
  \includegraphics[width=\textwidth]{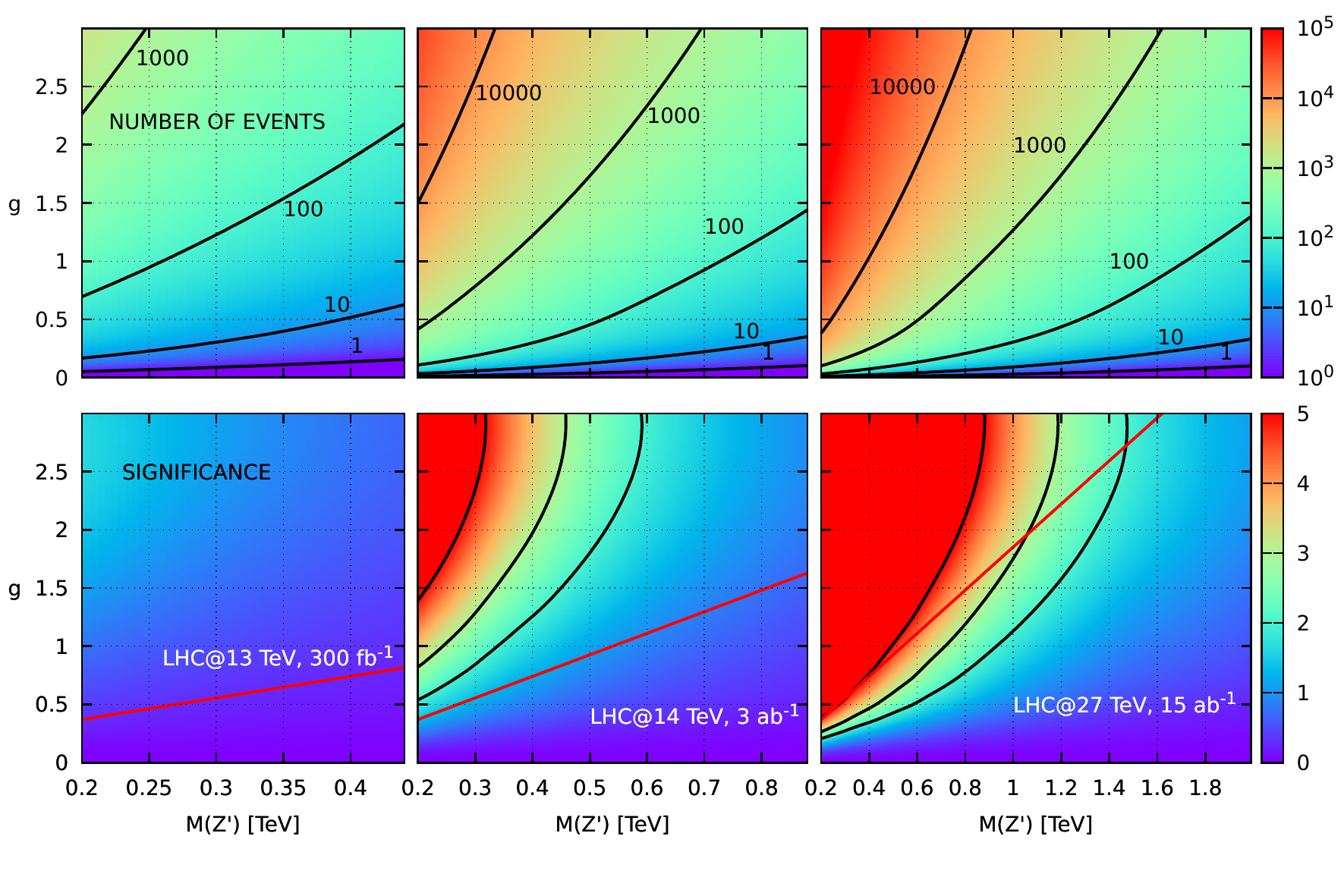}
  \caption{\label{fig:exclZP} Number of events (upper plots) and
    significance (lower plots) for the production of a $Z'$ coupled
    only to muons and taus, and decaying into muons, in the
    mass-coupling plane ($M_{\Zp},\,g$) in proton-proton collisions at
    13 TeV with $300\,{\rm fb}^{-1}$ (left), 14 TeV with $3\,{\rm
      ab}^{-1}$ (center) and 27 TeV with $15\,{\rm ab}^{-1}$ (right)
    of integrated luminosity. The contours corresponding to 1, 10,
    $10^2$, $10^3$ and $10^4$ events are shown in black in the upper
    plots, and the 2-, 3- and 5-sigma contours are shown in the lower
    plots.  The region above the red line, corresponding to
    $M_{\Zp}=g\times 540\,$GeV, is excluded by neutrino trident
    production.}
\end{figure}

As a second application of our lepton PDFs we consider here the
production of $Z'$-bosons. Here we make the very generic assumption
that we have a flavour diagonal $Z'$ that couples only to
leptons. Simple models that can account for that have been put forward
in the literature~\cite{He:1991qd}.  The resonance cross section is
given by
\begin{equation}
  \sigma(E)=\frac{2J+1}{(2s_1+1)(2s_2+1)} \frac{4\pi}{k^2}\frac{\Gamma^2}{4(E-M)^2+\Gamma^2} B_1B_2,
\end{equation}
where $J$ is the spin of the resonance, $s_1$ and $s_2$ are the spins of the incoming particles,
$k=E/2$,
$E$, $M$ and $\Gamma$ are the energy, mass and width of the resonance, and $B_1$, $B_2$ are the branching fractions of the resonance into
the initial and final state. In our case
\begin{equation}
  \sigma(E)= \frac{12\pi}{E^2}\left[\frac{\Gamma^2}{4(E-M)^2+\Gamma^2}\right] B_1B_2\,.
\end{equation}
In the narrow width limit we have
\begin{equation}
  \left[\frac{\Gamma^2}{4(E-M)^2+\Gamma^2}\right]\approx \pi \frac{\Gamma}{2} \delta(E-M).
\end{equation}
The hadronic cross section is then
\begin{equation}\label{eq:sigmaUsed}
  \sigma=  B_1B_2 \int \mathd \tau {\cal L}(\tau, s \tau)\frac{12\pi^2 \Gamma}{M} \delta(s\tau -M^2)
  = B_1B_2 {\cal L}\left(\frac{M^2}{s}, M\right) \frac{12\pi^2 \Gamma}{M s}\,.
\end{equation}
In the following we consider, for reference, a $Z'$ that couples
vectorially only to muons and taus (and to the corresponding
left-handed neutrinos), that is the least constrained in the model of
Ref.~\cite{He:1991qd}, and consider the $\mu^+\mu^-$ final state. In
this case, we have $B_1=B_2=1/3$, and the production proceeds from
both $\mu^+\mu^-$ and $\tau^+\tau^-$ annihilation.  The width is given
by
\begin{equation}
\Gamma=\frac{g^2}{4\pi} M\,.
\end{equation}
We now give a very rough estimate of the production rate and
significance of the corresponding signal, assuming an irreducible
Drell-Yan background.  For the muon energy resolution we interpolate
the measured points illustrated in Fig.~9 of
Ref.~\cite{Sirunyan:2018fpa}, and the reconstruction efficiency times
the acceptance was obtained by a rough interpolation of Fig.~2 of
Ref.~\cite{Sirunyan:2018owv}.  The significance plot for such an
object is shown in Fig.~\ref{fig:exclZP} (bottom panels), while the
upper panels show the number of expected events.
The plots were obtained in the following way. For each $Z'$ mass we
compute the number of signal events falling in a bin centered around
$M_{\Zp}$, with a size $b_w=\sqrt{\Gamma^2+r^2 M_{\Zp}^2}$, where $r$
is the muon $p_T$ resolution. We use formula~(\ref{eq:sigmaUsed}) for
the cross section, multiplied by a reduction factor $2
\arctan(b_w/\Gamma)/\pi$, to account for the loss on the sides of the
Lorentzian peak, and by the reconstruction efficiency times the
acceptance. We compute the Drell-Yan background using the code of
Ref.~\cite{Alioli:2008gx}, and integrate the cross section in the
given bin, multiplying also by the reconstruction efficiency times the
acceptance factor that we used for the signal.  The significance is
taken as the ratio of the number of signal events to the square root
of the number of background events.  Comparing our exclusion plot with
Fig.~2 of Ref.~\cite{Altmannshofer:2014pba} (for a direct limit see
\cite{Sirunyan:2018nnz}), we see that in the case of current and High
Luminosity LHC operation no relevant limit is found that is better
than the one coming from neutrino trident production of muon pairs,
that (according to
Ref.~\cite{Altmannshofer:2014pba,Altmannshofer:2016jzy}) yields
$M_{\Zp}/g\gtrsim 540\,$GeV, and is represented by the region above
the red line in the lower plots of Fig.~\ref{fig:exclZP}.  In the case
of the 27~TeV LHC, our method can yield exclusions of regions that are
still unexplored.

In the present study we have considered the Drell-Yan background as
irreducible. This may not be the case, since the hadronic activity
accompanying a leptonic collision is much smaller than the one
accompanying coloured parton collisions.  At the moment, we do not
know of any Parton Shower generator that can reliably generate the
initial state radiation from leptonic collisions, although there are
no serious obstacles to the implementation of such
effects~\cite{ThorbjornAndPeter}.  It is however easy to estimate
their size. The hardest radiation accompanying an initial state lepton
is another lepton, with average transverse momentum given by
\begin{equation}
  \langle \pt \rangle=\frac{\int_\Lambda^{M_{\Zp}} \frac{\mathd \pt}{\pt} \pt} {\int_\Lambda^{M_{\Zp}} \frac{\mathd \pt}{\pt}}
  = \frac{1}{\log \frac{M_{\Zp}}{\Lambda}} M_{\Zp}\,.
\end{equation}
A hadronic emission will happen only as next-to-hardest radiation, and
thus it will be suppressed by two powers of the logarithm.  We
emphasize also that in our case, roughly 50\%{} of the time the
hardest radiation will be a $\tau$, that may decay hadronically. The
tau will give rise to a very collimated jet. Thus, in around 50\%{}
plus $50\times 0.35$\%{} of the cases the hardest accompanying
radiation is leptonic, but also in the remaining fraction of hadronic
tau decays one may be able to reconstruct the tau by requiring a very
narrow jet. So, even if at the moment the effect of a jet or hadronic
veto on the efficiency of the signal selection cannot be estimated
reliably with a Monte Carlo, we have good reasons to believe that it
may yield a considerable reduction of the background.

\subsection{Doubly charged Higgs production}
From Fig.~\ref{fig:exclZP}, we see that it is the large Drell-Yan
background, rather than the lack of signal events, that limits the
reach for the detection of a $\Zp$. This suggests that we should turn
to signals that are essentially background free.  Opposite-sign
leptons of different flavours suffer for the presence of a large
$W^+W^-$ background, while same-sign leptons may be considered to be
essentially background free.

The production and decay of a doubly charged Higgs $H^{\pm\pm}$ via
lepton-lepton collisions may give a relevant signal for large enough
values of the coupling. Such an exotic particle usually arises in
extensions of the Standard Model which aim to accommodate neutrino
masses. For example, in the context of a type-II see-saw
mechanism~\cite{Magg:1980ut,Schechter:1980gr,Lazarides:1980nt,Mohapatra:1980yp},
a (non-renormalizable) dimension-5 operator added to the Standard
Model Lagrangian that can give rise to a Majorana mass term for the
neutrino, can be effectively generated by renormalizable interactions
with a triplet of scalar particles with ${\rm SU}(2)_L\times {\rm
  U}(1)_Y$ quantum numbers (3,2).  The triplet comprises a doubly
charged $H^{\pm\pm}$, a single charged $H^{\pm}$ and a neutral
component $H^{0}$.

The doubly charged state can couple both to leptons and to W
bosons. Therefore, its main production mechanisms are the pair
production via an s-channel intermediate $Z$ boson or photon, and the
associated production with a singly charged Higgs $H^{\pm}$. Various
experimental searches have been carried out by both the
ATLAS~\cite{Aaboud:2017qph} and CMS~\cite{Dev:2019ugu} collaborations
focusing on multi-lepton final states. Typically, these searches
assume a scenario in which the doubly charged Higgs decays
predominantly into leptons, assuming that the coupling to the W bosons
is negligible. The background is usually small since events with two
prompt and well isolated leptons with the same electric charge are
produced very rarely by Standard Model processes. A bump search is
performed for a narrow resonance in the same-sign lepton pair
invariant mass, which allows to put constraints on the lower value of
the mass of the doubly charged Higgs. The current limits exclude a
doubly charged Higgs with mass $M_{H^{\pm\pm}}\lesssim 800\,$GeV at
$95\%\,$CL.  These searches are insensitive to the coupling between
the $H^{\pm\pm}$ and the leptons, and only the leptonic branching
ratios matter, provided the coupling is large enough for the decay to
occur inside the fiducial volume of the detector.

\begin{figure}
  \centering
  \includegraphics[width=\textwidth]{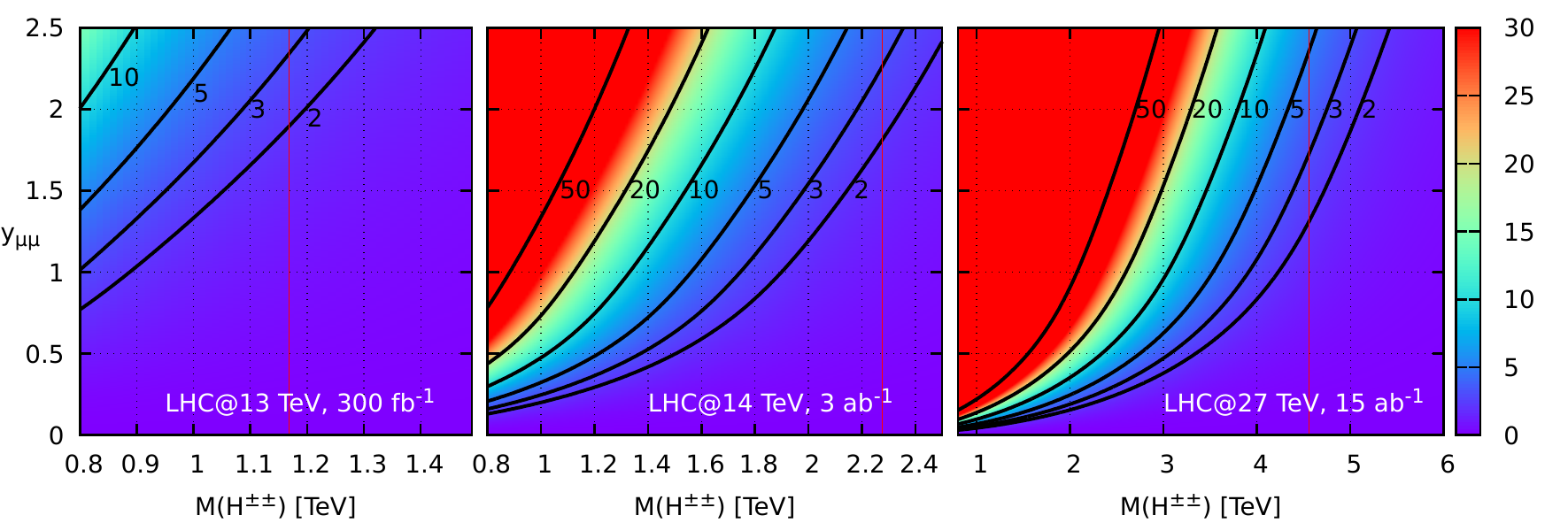}
  \caption{\label{fig:exclHpp} Number of events for the productions of
    doubly charged Higgs production.  The solid line correspond to 2,
    3, 5, 10, 20 and 50 events. The thin red lines represent the
    projected limits from using $H^{\pm\pm}$-pair
    production~\cite{deMelo:2019asm} at 139~fb$^{-1}$, 3~ab$^{-1}$ and
    15~ab$^{-1}$ for the current LHC, the High Luminosity and the High
    Energy upgrades respectively.}
\end{figure}

We consider here the direct resonant production of a single
$H^{\pm\pm}$ from lepton-lepton annihilation, whose rate is
proportional to the square of the $y_{l_1l_2}$ Yukawa coupling. This
is complementary to the searches mentioned earlier. We observe that
this search strategy for the signal is analogous to the previous study
on the $\Zp$, so that we can effectively apply the same procedure. At
variance with that case, the Standard Model background is drastically
reduced, and we assume here that the process is essentially background
free. The signal signature is indeed given by a pair of same-sign
leptons with no missing energy and very limited activity in the event.
We consider, for concreteness, a $H^{\pm\pm}$ that couples only to
muons.  In Fig.~\ref{fig:exclHpp}
we plot the number of detected events in the $[M(H^{\pm\pm})$,
  $y_{\mu\mu}]$-plane.  We don't consider the mass region below $0.8$
TeV, that has already been ruled out by ATLAS and
CMS~\cite{Aaboud:2017qph,Dev:2019ugu}.  Projections to higher
luminosities of these analyses have been given in
Ref.~\cite{deMelo:2019asm}, and are reported in the figure. From the
plot, we see that at the present LHC with 300~fb$^{-1}$ a few events
will be available for masses above the projected exclusion limit of
Ref.~\cite{deMelo:2019asm}, that corresponds to 1.168 TeV at
139$\,{\rm fb}^{-1}$.  The same is true at the High Luminosity LHC,
where the projected limit is 2.276~TeV, and at the High Energy LHC,
where the projected limit is 4.56~TeV. We thus conclude that, for
sufficiently large coupling, the $s$-channel production of a doubly
charged Higgs may have a mass reach comparable to analyses relying
upon pair production.

\subsection{Single leptoquark searches}

Leptoquarks (LQs) are hypothetical particles which couple a quark to a
lepton at the tree level. They can be either scalar or vector fields
and are coloured under the Standard Model $SU(3)$ colour group. They
arise in several extensions of the Standard Model and provide an
appealing explanation for tensions in flavour physics.  For a review
of the various aspects of the LQ physics we refer to the
Refs.~\cite{Davidson:1993qk,Hewett:1997ce,Nath:2006ut,Dorsner:2016wpm}.

As an illustrative model, we consider a chiral $R_2$ LQ of charge
$5/3$ (following the nomenclature in Ref.~\cite{BUCHMULLER1987442}),
which couples to the conjugate of the left charged leptons to right
u-type quarks. A summary of LQ searches based on both single and pair
LQ production can be found in~\cite{Schmaltz:2018nls}. According to
this analysis, based on $36\,$fb$^{-1}$ data at $13\,$TeV, the point
$m_{LQ}=2\,$TeV, $y_{eu}=0.3$ in the mass-Yukawa coupling parameter
space is still allowed for a LQ which couples only to the quark up and
the positron\footnote{It is not excluded even if one considers the
  more stringent bounds coming from the recast of the experimental
  results on the measurement of the weak charge in atomic systems (see
  Appendix B in Ref.~\cite{Schmaltz:2018nls}).}. For the case of a LQ
which couples only to the quark up and the muon and for the same value
of the LQ mass, slightly larger couplings remain unconstrained, as for
example $y_{\mu u}=0.5$. In the following, we consider these two
points as our benchmark scenarios. As for the width, we assume it is
dominated by the 2-body decay and it is given by
\begin{equation}
\Gamma_{LQ} = \frac{y_{lq}^2 }{16 \pi}m_{LQ},
\end{equation}
neglecting all fermion masses.

Leptoquarks can be searched for via the $s$ channel
process $\ell+q \to \ell+q$, where both the lepton and the quark arise
as partons in the proton beams.
Having at our disposal a precise determination of the lepton
densities, we can investigate the sensitivity reach of this production
mechanism.\footnote{This process was also considered in ref.~\cite{Ohnemus:1994xf},
based upon a simple estimate of the lepton pdf.}
  Here, we consider the
two subprocesses $e^+ + u \to e^+ + u$ and $\mu^+ + u \to \mu^+ + u$
separately. As for the background, we assume as main source the
associated production of a jet and a $W$ boson decaying
leptonically. We require that the lepton and the jet are both central,
$|\eta|<2.5$, and with transverse-momentum larger than
$500\,$GeV. Furthermore, since the signal is a lepton+jet system
balanced in the transverse plane, for the background
estimate we veto missing transverse-momentum
associated to the neutrino larger than $50\,$GeV.

\begin{figure}
  \centering
  \includegraphics[width=0.49\linewidth]{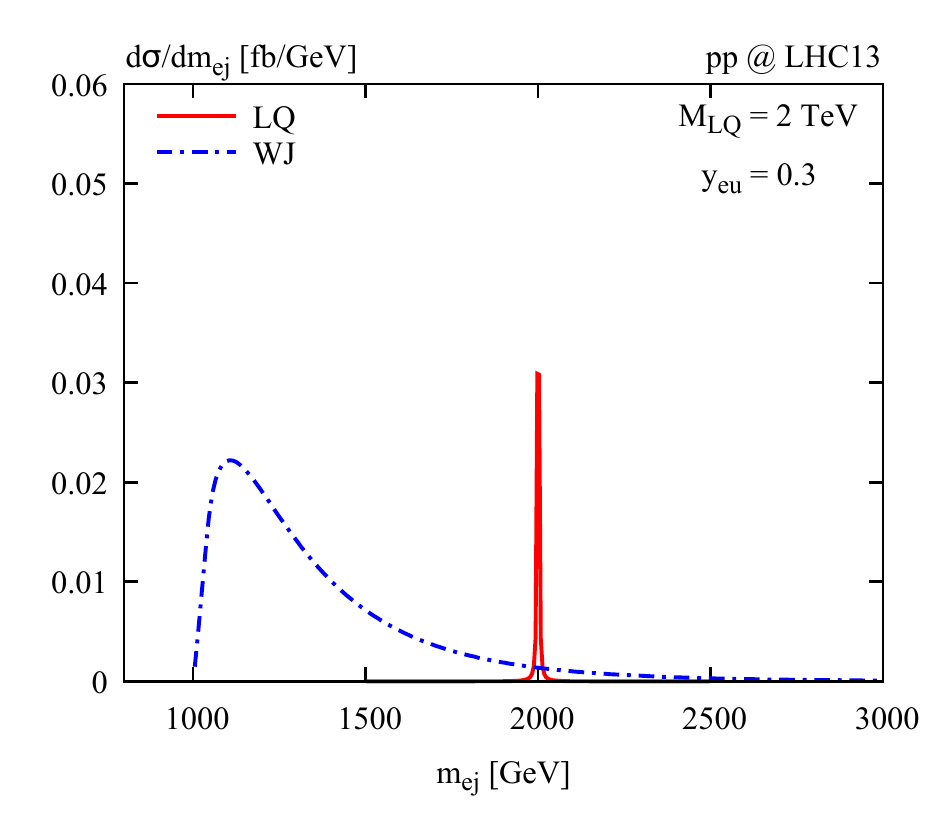}
  \includegraphics[width=0.49\linewidth]{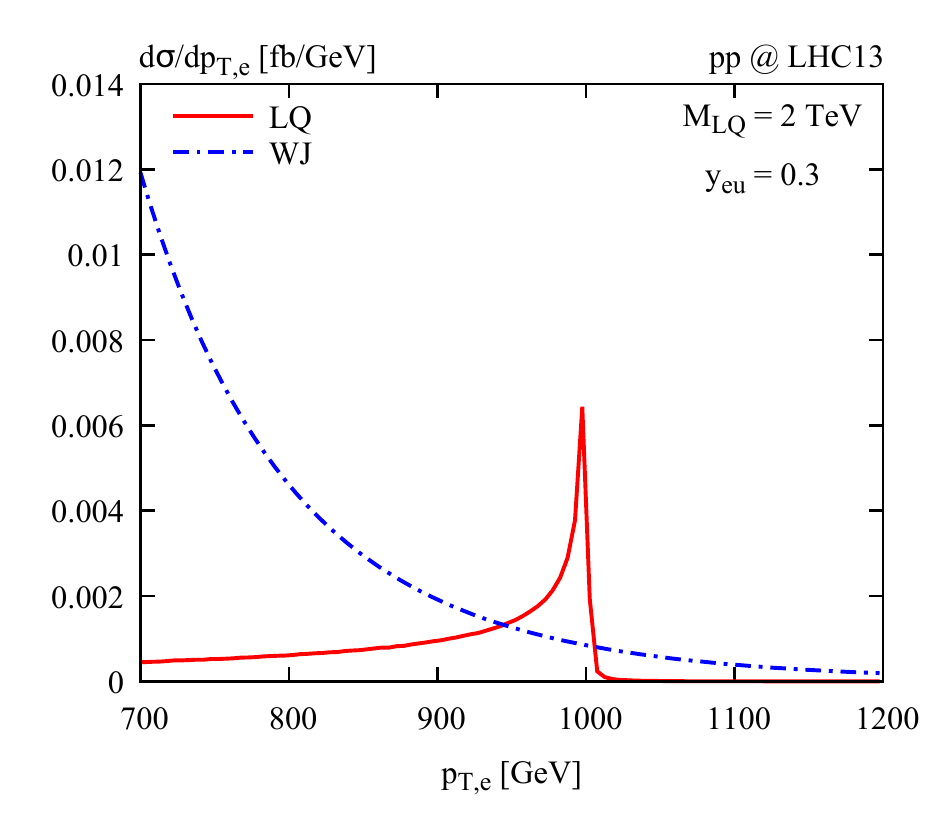}
  \caption{\label{fig:LQ_eu} Invariant mass of the lepton plus jet system (left) and electron transverse-momentum (right) distributions. The signal (red line) is due to a LQ of mass $M_{LQ}=2\,$TeV and Yukawa coupling $y_{eu}=0.3$ in the hypothesis of minimal coupling of the LQ between first-generation quarks and leptons. $W(l\nu)+1\,$jet background is shown in blue. The selection cuts are $p_{T,l},p_{T,j}>500\,$GeV, $|\eta_{l}|,|\eta_{j}|<2.5$ and $p_{T,\text{miss}} < 50\,$ GeV.}
\end{figure}

\begin{figure}
  \centering
  \includegraphics[width=0.49\linewidth]{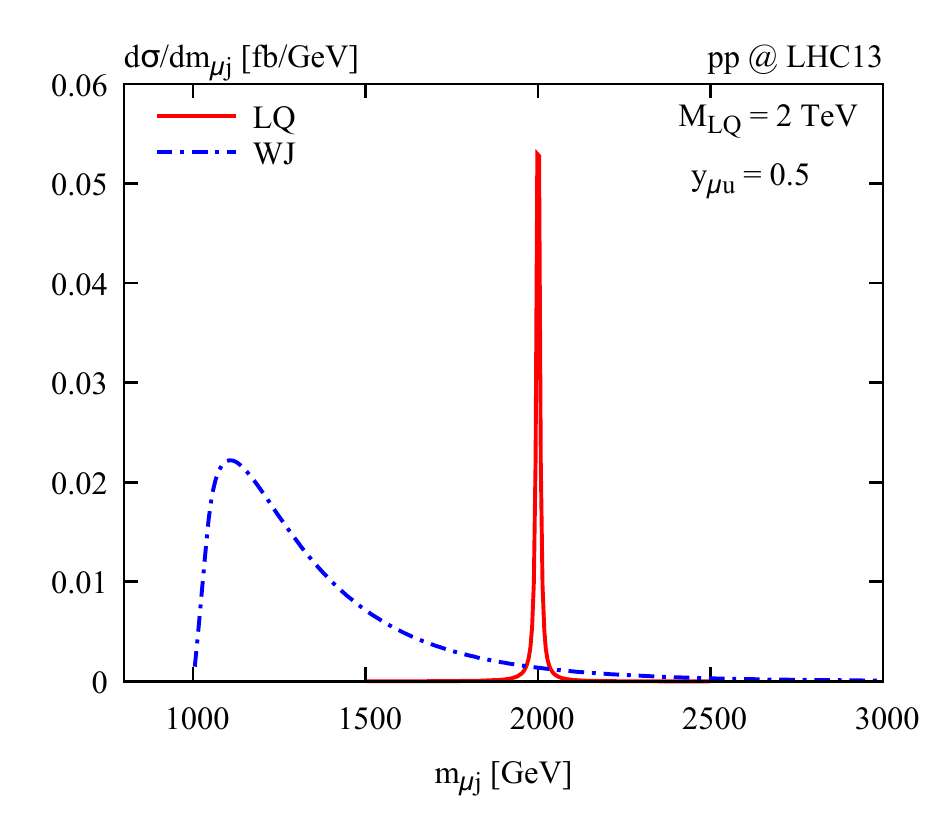}
  \includegraphics[width=0.49\linewidth]{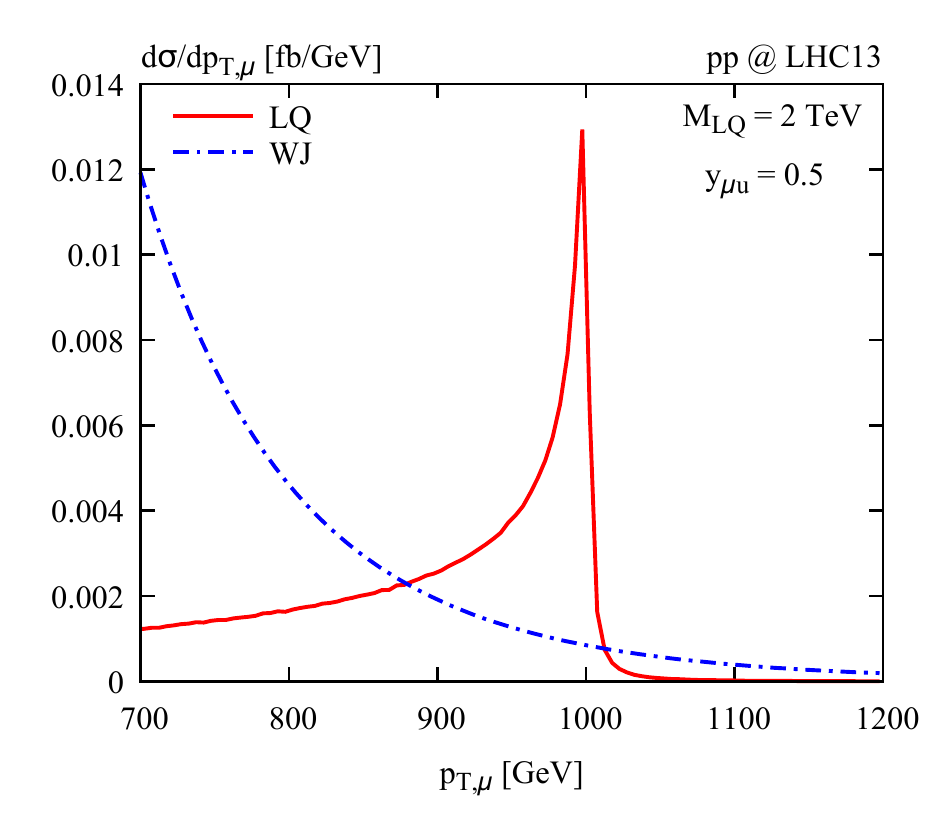}
  \caption{\label{fig:LQ_muu} Invariant mass (left) and muon
    transverse-momentum (right) distributions of the muon+jet
    system. The signal (red line) is due to a LQ of mass
    $M_{LQ}=2\,$TeV and Yukawa coupling $y_{LQ}=0.5$ in the hypothesis
    of minimal coupling of the LQ between first-generation quarks and
    second-generation leptons. $W(l\nu)+1\,$jet background is shown in
    blue. Selection cuts: $p_{T,\ell},p_{T,j}>500\,$GeV,
    $|\eta_{\ell}|,|\eta_{j}|<2.5$ and $p_{T,\text{miss}} < 50\,$
    GeV.}
\end{figure}

In Figs.~\ref{fig:LQ_eu} and~\ref{fig:LQ_muu}
we plot the invariant mass of the lepton+jet system and the charged
lepton transverse-momentum distributions produced in $pp$ collisions
at $13\,$TeV for the positron-up and the antimuon-up processes
respectively. A good sensitivity to the LQ is reached with clear peaks
in both distributions.

For a more quantitative and immediate comparison to Ref.~\cite{Schmaltz:2018nls}, in tab.~\ref{tab:LQ_res}
\begin{table}
  \centering
  \begin{tabular}{|c|c|c|}
    \hline
    & $\sigma$ [fb] & \#events\\
    \hline
    $e^+ + u$ $(m_{LQ}=2\,\text{TeV},y_{eu}=0.3)$  & $0.40$ & $14$\\
    \hline
    $\mu^+ + u$ $(m_{LQ}=2\,\text{TeV},y_{\mu u}=0.5)$& $1.07$ & $36$\\
    \hline
    $W^++j$                                     & $0.14$ & $5$\\
    \hline
  \end{tabular} \caption{\label{tab:LQ_res} Cross sections and number of expected events for an integrated luminosity of $36\,$fb$^{-1}$ in the lepton+jet mass window $1950\,\text{GeV}< m_{\ell j} <2050\,$GeV. These numbers do not include the charge conjugate process, that is negligible.}
\end{table}
we report the cross sections and the number of expected events in the
lepton+jet mass window $1950\,\text{GeV}< m_{\ell j} <2050\,$GeV
corresponding to an integrated luminosity of $36\,$fb$^{-1}$. Our
finding is that the signal-to-background ratio $S/\sqrt{B}$ is very
good, about $6.2$ and $16$ for the electron-up and the muon-up
processes respectively. From this preliminary analysis, the single
$s$-channel LQ mechanism outlined above seems able to reach regions of
the parameter space that cannot be accessed with current methods.

\section{Conclusions}
\label{sec:conclu}

In this work we have carried out a computation of the lepton densities
in the proton, up to NLO accuracy. The computation relies upon the good
quality of the available data on the electroproduction structure functions
and the proton form factors, and is carried out in full analogy with what was
done for the photon density in
Refs.~\cite{Manohar:2016nzj,Manohar:2017eqh}. As in the photon case,
the inclusion of NLO corrections is mandatory in order to reach an
accuracy at the few percent level.  It is in principle possible to
increase the precision of our lepton PDFs up to the NNLO-QCD level, as was
done in Refs.~\cite{Manohar:2017eqh} for the photon.

We have used our result to extend the pdf set {\tt
  NNPDF31\_nlo\_as\_0118\_luxqed} of Ref.~\cite{Bertone:2017bme} with
the inclusion of leptons. The set so obtained will be soon made
available in the LHAPDF~\cite{Buckley:2014ana} format, under the name
{\tt LUXlep-NNPDF31\_nlo\_as\_0118\_luxqed}.

With respect to the corresponding calculation in the photon case, some
novel requirements have emerged for the lepton PDFs that were not
present there. These are better understood if we remind that we have
adopted the following rules for the determination of the perturbative
order of the calculation: quark densities are of order 1 (actually of
order $(\as(\mu) L)^n$, with $L=\log(\mu/\Lambda)$ and $\Lambda$ a
typical hadronic scale for all $n$); photon densities are of order
$\aem L$ relative to the quark densities; and lepton densities are of
order $\aem L$ relative to the photon densities, and thus of order
$\aem^2 L^2$ relative to the quark densities. NLO accuracy requires
that terms of order $\aem$ should be retained for the photons, and
terms of order $\aem^2 L$ should be retained for the leptons, thus
adopting the criterion $ L \approx \as^{-1}(\mu)$. According to this
counting, leading order electromagnetic plus mixed QED-QCD (of order
$\aem\as$) splitting functions (computed in \cite{deFlorian:2015ujt})
are all what
is needed to maintain NLO
accuracy of the photon density, while in the case of lepton densities
a term of order $\aem^2$ (computed in \cite{deFlorian:2016gvk})
is also needed.  In fact on the right hand
side of the evolution equation for the lepton densities we expect
terms of the form
\begin{equation}
  \frac{\partial f_l}{\log \mu^2}=p_{l\gamma} \otimes f_\gamma+p_{l q}\otimes f_q,
\end{equation}
where in the first term we have contributions of order $\aem^2 L$ and
$\aem^2$ from the leading and NLO contributions to $f_\gamma$, and the
second term, of order $\aem^2$, needs the inclusion of the $\aem^2$
splitting function $p_{l q}$.  This splitting function is not normally
included in the electromagnetic evolution of parton densities, but is
needed in our case, and we implemented it in \hoppet{} in order to
complete our work.  Another aspect is the treatment of higher-order
electromagnetic terms, giving rise to contributions of order $\aem^3
L^3$ to the lepton PDFs. If we adopt the commonly used rule that $\aem
\approx \as^2$, and thus $\aem L^2\approx 1$, these terms should be
counted as NLO in our calculation.  In the framework of the photon
PDFs, similar consideration lead to the inclusion of terms of order
$\aem^2 L^2$, that in that case arise only from the electromagnetic
charge renormalization.  In the lepton PDF case, the $\aem^3 L^3$
terms arise from collinear radiation of photons from the final lepton.
We have devised different methods to include these terms in our
calculation.

Our calculation can be extended to NNLO-QCD level, i.e. including all
terms of ${\cal O}(\aem^2)$. We can expect a reduction in the error
due to higher-order corrections of the same order as what was found
for the photon PDF, i.e. by a factor 2-3 (see Fig.~13 in LUX2). Since
the error due to missing higher-order effects dominates in most of the
$x$-range, this would yield a substantial increase in precision. While
it is important to know that our calculation can be further improved,
at present we see no compelling reason to target this level of
precision.

The inclusion of the lepton parton densities in the proton adds new
production mechanisms for Standard Model and New Physics processes at
the LHC. Lepton-lepton electromagnetic scattering can give rise to
final states with different flavour and/or same sign leptons. We found
that these phenomena may be observable at the LHC. Besides being
useful to test the underlying theory, these processes may also be used
to assess the structure of the underlying event in lepton-initiated
processes. In view of the large contribution of the elastic component
to the lepton PDF, there must be a substantial fraction of events with
large rapidity gaps, containing only the matching leptons of opposite
charge arising from the photon splitting process. Even if we let aside
the possible presence of rapidity gaps, we know that the hadronic
activity in lepton-initiated processes must be greatly reduced.  At
present, to our knowledge, there are no shower Monte Carlo that can
simulate lepton-initiated processes, although we may assume that they
will become available in the near future. It is likely that when these
tools will become available, it will be possible to optimize methods
for rejecting processes initiated by coloured partons (with respect to
those initiated by leptons) based upon the accompanying event
activity, thereby reducing potential backgrounds to searches targeting
lepton-initiated New Physics processes.

In this work we have also considered few applications of the lepton
PDFs to basic Standard Model scattering processes and to some selected
New Physics processes. We have found that lepton scattering at low
transverse momentum (above 20 GeV) is likely to be observable at the
LHC, with rates increasing from a handful of events with the current
LHC settings, up to several hundreds for the High Luminosity LHC, and
few thousands for a High-Energy LHC.

As example of searches of New Physics, we have considered the case of
leptonic production of a hadrophobic $\Zp$ that couples only to muons
and taus.  By considering the $\mu^+\mu^-$ final states, we have found
non-negligible production rates, and a relevant significance over the
Drell-Yan background in a large region of the mass-coupling plane,
although, for the case at hand, neutrino scattering data already
excludes a large fraction of this region.

Production of doubly charged resonances coupled to leptons can yield a
signal that is essentially background free, consisting of same-sign
leptons. We have considered the production of a doubly charged Higgs
that couples only to muons.  This object can also be pair-produced, or
produced in association with a singly charged partner at the LHC, and
in this case the signal does not depend upon its coupling to leptons,
but only upon its branching ratios. With the current LHC and its
High-Luminosity and high-energy upgrades we find that, for
sufficiently large couplings, the leptonic production can reach the
limits that can be set using pair production.

Incoming leptons colliding with quarks can give rise to $s$-channel
production of leptoquarks. In this case, the rates are higher with
respect to the lepton collision processes, since they benefit from the
valence density of the quarks.  We have considered two benchmark
points for the production of a scalar leptoquark coupled either to up
quarks and positrons or to up quarks and anti-muons, with couplings
$h_{ue}=0.3$ for the first case and $h_{u\mu}=0.5$ for the second
case, which are currently not excluded.  In both cases we observe a
relevant signal over the $W^+ j$ background, both in the invariant
mass of the lepton-jet final state and in the lepton spectrum. This
production mechanism is thus very promising, and further investigation
in this direction are undergoing~\cite{Uli}.

The applications that we have reported here only serve as examples of
what could be achieved in the framework of lepton-initiated
processes. We remark again that this framework is quite new, and in
order to develop it further, extensions of Monte Carlo generators that
can handle lepton densities in the proton are needed.  Likewise, NLO
calculations of lepton-initiated processes should be performed, and
implemented in NLO+PS frameworks. As stressed several times in this
work, these NLO corrections are of the same order as the typical NLO
corrections in hadronic collisions, i.e. from 1 to few 10\%{}, and
thus they are necessary in order to achieve a reasonable accuracy.
Interfacing them to parton showers is also necessary in order to
understand to what extent vetoing over hadronic activity affects
lepton-initiated processes, and thus can be used to limit potential
backgrounds.  In the studies that we have performed, we have found
that lepton-initiated processes in proton collisions have the
potential to increase the reach of New Physics searches at the LHC,
thus justifying the theoretical effort needed for their study.

\section*{Acknowledgments}
L.B. and G.Z. thank Milano Bicocca for hospitality while part of this
work was carried out.  We thank Gavin Salam and Uli Haisch for several
inspiring discussions and input on the manuscript.  Furthermore we are
grateful to Wojtek Bizon, Marta Calvi, Stefano Carrazza, Simone
Gennai, Gino Isidori, Peter Richardson, Torbj\"orn Sj\"ostrand and
Andrea Wulzer for useful exchanges. This work is supported in part by
the Swiss National Foundation under contracts 200020\_188464 and
IZSAZ2\_173357. P.N. acknowledges support from Fondazione Cariplo and
Regione Lombardia, grant 2017-2070, and from INFN.

\appendix

\section{Partonic calculation}
\label{app:partoniccalc}
We need to compute the process shown in Fig.~\ref{fig:basicProcess} (right),
where now $q=xp$ and $q^2=0$.
We have
\begin{eqnarray*}
  r \cdot q & = & \frac{\ecm^2}{2},\\
  q \cdot k & = & \frac{\ecm^2 + M^2}{4} - \frac{\ecm^2 -
  M^2}{4} y\\
  r \cdot k & = & \frac{\ecm^2 + M^2}{4} + \frac{\ecm^2 -
  M^2}{4} y,
\end{eqnarray*}
where $y$ is the cosine of the scattering angle in the photon-scalar CM frame.
The squared amplitude can be written as
\begin{eqnarray*}
  | A_r |^2 & = & \frac{g^2}{2} \Bigg\{ \frac{\tmop{Tr} \left[ \left( \cancel{k}
  + M \right) \left( \cancel{k} - \cancel{r} \right) \gamma^{\mu} \left( -
  \cancel{\bar{k}} \right) \gamma_{\mu} \left( \cancel{k} - \cancel{r} \right)
  \right]}{(k - r)^4} \\
  &  & + \frac{\tmop{Tr} \left[ \left( \cancel{k} - M \right) \gamma^{\mu}
  \left( \cancel{k} - \cancel{q} + M \right) \left( - \cancel{\bar{k}} \right) \left(
  \cancel{k} - \cancel{q} + M \right) \gamma_{\mu} \right]}{((k - r)^2 - M^2)^2}\\
  &  & + 2  \frac{\tmop{Tr} \left[ \left( \cancel{k} - M \right)
  \gamma^{\mu} \left( \cancel{k} - \cancel{q} + M \right) \left( - \cancel{\bar{k}}
  \right) \gamma_{\mu} \left( \cancel{k} - \cancel{r} \right) \right]}{(k - r)^2 
  ((k - r)^2 - M^2)} \Bigg\},
\end{eqnarray*}
where the overall $1 / 2$ factor is the spin average for the photon, and two
minus signs, one for the fermion loop and one for the photon spin projection,
compensate each other. The coupling constant for the scalar has been omitted,
since it cancels when dividing by the Born cross section. The real phase space
is
\begin{equation}
 \frac{1 - \frac{M^2}{\ecm^2}}{16 \pi} \mathd y,
\end{equation}
and the (partonic) flux factor is
\begin{equation}
\frac{1}{4 q \cdot r} = \frac{1}{2 \ecm^2} .
\end{equation}
The Born cross section is
\begin{equation}
\sigma_B \delta (\ecm^2 - M^2) = \frac{1}{2}  \frac{1}{2 M^2}  (2
   M^2) 2 \pi \delta (\ecm^2 - M^2)=\pi\, \delta (\ecm^2 - M^2),
\end{equation}
where the first factor is the spin average, the second factor is the flux
factor, the third factor is the result of the trace
\begin{equation}
\tmop{Tr} \left[ \left( \cancel{r} + z\right)\left( \cancel{q} + M \right) z \cancel{q} \right]=2 M^2,
\end{equation}
and the $2 \pi \delta (\ecm^2 - M^2)$ is the single particle phase
space.

The real cross section $\sigma_r$ is computed in four dimensions by
standard means, multiplying the square amplitude, the flux factor and the
phase space. It has the form
\begin{equation}
  \frac{\sigma_r}{\sigma_B} = \int \mathd y \left[ A (y, z) +
    \frac{1}{1 + y} B (z) \right],
\end{equation}
where $z = M^2 / \ecm^2$

According to the FKS prescription~\cite{Frixione:1995ms}, the parton model result is
\begin{eqnarray*}
  \frac{\sigma_r}{\sigma_B} & = & \int_{- 1}^1 \mathd y \left\{ A (y, z) +
  \left( \frac{1}{1 + y} \right)_+ B (z) \right\}\\
  & + & \int \mathd z \frac{\aem}{2 \pi} \left\{ (1 - z) \Plgamma (z)
  \left[ \frac{1}{1 - z} \log \frac{M^2}{z \mu_F^2} + 2 \frac{\log (1 - z)}{1
  - z} \right] + 2 z (1 - z) \right\} \delta (\ecm^2 z - M^2)\\
  & = & \int_{- 1}^1 \mathd y \left\{ A (y, z) + \left( \frac{1}{1 + y}
  \right)_+ B (z) \right\}\\
  & + & \frac{\aem}{2 \pi}  \frac{1}{M^2} \left\{ z (1 - z) \Plgamma
  (z) \left[ \frac{1}{1 - z} \log \frac{M^2}{z \mu_F^2} + 2 \frac{\log (1 -
  z)}{1 - z} \right] + 2 z^2 (1 - z) \right\}
\end{eqnarray*}
where the second term is the collinear remnant (see Eq.(2.102) of ref~\cite{Frixione:2007vw}).
We get
\begin{equation}
  \int_{- 1}^1 \mathd y \left\{ A (y, z) + \left( \frac{1}{1 + y} \right)_+ B
   (z) \right\} = \frac{\aem}{2 \pi}  \frac{1}{M^2} \left\{ z \Plgamma
   (z) \log \left( \frac{1}{z} \right) + 2 z^2 (1 - z) \right\}
\end{equation}
Thus
\begin{eqnarray*}
  \frac{\sigma_r}{\sigma_B} & = & \frac{\aem}{2 \pi}  \frac{1}{M^2} \left\{ z
  \Plgamma (z) \left[ \log \frac{M^2}{\mu_F^2} + \log \frac{(1 -
  z)^2}{z^2} \right] + 4 z^2 (1 - z) \right\}.
\end{eqnarray*}
The full parton model formula at NLO is thus
\begin{eqnarray}
  \frac{\sigma}{\sigma_B} & = & \int \mathd x \fl (x, \mu_F^2) \delta (S x -
  M^2) \nonumber\\
  & + & \frac{\aem}{2 \pi}  \frac{1}{M^2} \int \mathd x f_{\gamma}(x,  \mu_F^2) 
  \left\{ z \Plgamma (z) \left[ \log \frac{M^2}{\mu_F^2} + \log \frac{(1 -
  z)^2}{z^2} \right] + 4 z^2 (1 - z) \right\}, \phantom{aa}
\end{eqnarray}
where $z = M^2 / \ecm^2 = M^2 / (S x)$.

\section{The effect of the lepton mass}\label{app:masseffects}
There are cases when the effect of the lepton mass on the lepton PDF
cannot be neglected. For the muon, for instance, it turns out that the
smallest values of $Q^2$ that contribute to the lepton parton density
is of the same order of the muon mass squared, and for the tau there
is a contribution from a range of $Q^2$ values below the tau mass
squared.  For the electron, one would be inclined to believe that the
mass should not matter. However, we recall that, since the proton has
an overall electric charge, at very high energy it carries an
accompanying electromagnetic field that can be described as a
superposition of virtual photons, that in turn can materialize into
pairs of nearly massless leptons.  This implies that for very small
$x$ the lepton PDF for a truly massless lepton should diverge, due to
the elastic contribution.

It is not difficult to carry out the computation of the diagrams of Fig.~\ref{fig:basicProcess}~(a)
with a finite lepton mass. We obtained
\begin{eqnarray}
  \sigma &=& \frac{\pi}{M^2} \left(\frac{\aem}{2\pi}\right)^2
             \int_\frac{M^2}{S}^{\frac{M^2}{(M+\ml{})^2}} \mathd
\zl \int_x^1 \frac{dz}{z} \int_{\frac{\mpr^2
             x^2}{1-z}}^{\frac{\ecmsq(1-z)}{z}} \frac{dQ^2}{Q^2}
\nonumber  \\ && \Bigg\{
\Plgamma(\zl) \left[ F_2 \left(z \Pgammaq(z) + \frac{2 \mpr^2
      x^2}{Q^2}\right)
  - F_L z^2\right]\log\frac{M^2(1-\zl)}{\zl^3 \left(Q^2+\frac{\ml^2}{\zl(1-\zl)}\right)}
\nonumber   \\ &&  + F_2 \left[4(z-2)^2\zl(1-\zl)-z \Pgammaq(z) \right]
\nonumber \\ &&   +F_Lz^2 \Plgamma(\zl)-\frac{2\mpr^2 x^2}{Q^2}F_2
\nonumber   \\ &&
+ \frac{\ml^2F_2}{\ml^2+Q^2\zl(1-\zl)} 
                  \left[ z\Pgammaq(z) -8\zl(1-\zl)\left(1-z-\frac{\mpr^2x^2}{Q^2}\right)
             +\frac{2\mpr^2 x^2}{Q^2}    \right] \nonumber \\
         &&- \frac{\ml^2F_Lz^2}{\ml^2+Q^2\zl(1-\zl)}\left[2- \Plgamma(\zl)\right]
          \Bigg\}\,.
\label{eq:sigmafinLepMass}
\end{eqnarray}
For lepton masses far below the proton mass, the last line is suppressed
by powers of $\ml/\mpr$. As far as the diagrams of Fig.~\ref{fig:basicProcess}~(b) are
concerned, for lepton masses below the scale we are probing (i.e. the
mass $M$ of the heavy fermion) the mass of the physical lepton yields
only power suppressed effects, and can be safely neglected.  We thus
notice that the only lepton mass effect that is relevant to our result
is the modification of the argument of the logarithm, and the addition of the
last line in eq.~(\ref{eq:sigmafinLepMass}).

\section{Simplification of the integrand} \label{app:twoDimIntegrand}
It is convenient to rewrite formula~(\ref{eq:leptonpdf1}) as an
integral in $Q^2$, $\xbj$ and a third variable.  At fixed $Q^2$,
$\xbj$, the dependence of formula~(\ref{eq:leptonpdf1}) upon the third
variable is analytical, and its integration can be performed with
algebraic means. We rewrite the integration as
\begin{eqnarray}
  &&
  \int_{\xl{}}^1 \frac{\mathd x}{x}  \int_x^1 \frac{\mathd z}{z}
   \int_{\frac{\mpr^2 x^2}{1 - z}}^{M^2 (z)} \frac{\mathd Q^2}{Q^2}= \nonumber \\
  &  &\phantom{aaaaaa}
  \int_{\xl{}}^1 \frac{\mathd \xbj{}}{\xbj{}}
  \int_0^{\infty} \frac{\mathd Q^2}{Q^2} \int_{\xi}^1
  \frac{\mathd z}{z} \left[ \theta (M^2_z - Q^2) - \theta \left( \frac{\mpr^2
  \xbj^2 z^2}{1 - z} - Q^2 \right) \right] ,
\end{eqnarray}
where we have used the identities $\xbj=x/z$ and $\xi=\xl/\xbj$.
We have the freedom to consider the $Q^2$ integration as an oriented one,
since in all cases the region where the upper bound is below the lower bound
is very small.
In the case $M^2 (z) = \muF^2 / (1 - z)$, under the safe assumption that $\muF>\mpr$,
the upper limit is always above the lower limit. We must turn the $\theta$ functions
into limits on the $z$ integration. We find
\begin{equation}
  1 - z < \frac{\muF^2}{Q^2},\quad\quad \mpr^2 \xbj^2 z^2 - (1 - z) Q^2 > 0 \label{eq:ineqz12}
\end{equation}
for the first and second theta function in the square bracket respectively.
In the second inequality the equal sign holds when
\begin{equation} z_{\pm} = \frac{Q^2}{2 \mpr^2 \xbj^2} \left[ \pm \sqrt{1 +
   \frac{4 \mpr^2 \xbj^2}{Q^2}} - 1 \right]\,.
\end{equation}
The two solutions have opposite signs, and the second inequality in~(\ref{eq:ineqz12}) holds if
\begin{equation}
  z>z_+=\frac{Q^2}{2 \mpr^2 \xbj^2} \left[ \sqrt{1 +
   \frac{4 \mpr^2 \xbj^2}{Q^2}} - 1 \right]=\frac{2}{1+\sqrt{1 + \frac{4 \mpr^2
   \xbj^2}{Q^2}}}\,.
\end{equation}
Thus, the term proportional to $\theta (M_z^2 - Q^2)$ has the $z$ limits
\begin{equation} 1 > z > z_{\min}^{(1)} \equiv \max \left( \xi, 1 -
   \frac{\muF^2}{Q^2} \right),
\end{equation}
where $\xi=\xl/\xbj$.
For the term proportional to the second $\theta$ we have
\begin{equation} 1 > z > z_{\min}^{(2)} \equiv \max \left(\xi, \frac{2}{1+\sqrt{1 + \frac{4 \mpr^2
   \xbj^2}{Q^2}}}\right)\,.
\end{equation}
Under the safe assumption that $\muF>\mpr$ we can easily verify that
\begin{equation}
  1-\frac{\muF^2}{Q^2}<\frac{2}{1+\sqrt{1+\frac{4\mpr^2\xbj^2}{Q^2}}},
\end{equation}
that in turn implies $z^{(1)}_{\min}\leq z^{(2)}_{\min}$. 
Thus, the $z$ integral becomes
\begin{equation} \int_{z^{(1)}_{\min}}^1 \frac{\mathd z}{z} - \int_{z^{(2)}_{\min}}^1
  \frac{\mathd z}{z}=\int_{z^{(1)}_{\min}}^{z^{(2)}_{\min}} \frac{\mathd z}{z}
  \,.\label{eq:zintdiff}
 \end{equation}
 The $Q^2$ integration is implicitly limited to values of $Q$ such
 that $z^{(1)}_{\min} \neq z^{(2)}_{\min}$. The equal sign holds if
 the following inequalities
\begin{eqnarray}
  \xi & > 1 - \frac{\muF^2}{Q^2}\phantom{aaaaa} & \Rightarrow Q^2 <
  \frac{\muF^2}{1 - \xi}\,,\\
  \xi & >  \frac{2}{1+\sqrt{1 + \frac{4 \mpr^2
   \xbj^2}{Q^2}}}
                                                          & \Rightarrow
            Q^2  <  \frac{\mpr^2 \xl^2}{1 - \xi}  . \label{eq:condq2z2}
\end{eqnarray} 
hold at the same time.
Thus, if
\begin{equation} Q^2 < \min \left( \frac{ \mpr^2 \xl^2}{1 - \xi},
   \frac{\muF^2}{1 - \xi} \right) = \frac{\xl^2 \mpr^2}{1 - \xi}
\end{equation}
(where we have assumed $\muF>\mpr$)
the integral vanishes, and the right-hand side of the above equation
is the $Q^2$ lower limit. So, our original integration is rewritten as
\begin{equation}
   \int_{\xl{}}^1 \frac{\mathd x}{x}  \int_x^1 \frac{\mathd z}{z}
   \int_{\frac{\mpr^2 x^2}{1 - z}}^{M^2 (z)} \frac{\mathd Q^2}{Q^2}=
   \int_{\xl{}}^1 \frac{\mathd \xbj{}}{\xbj{}}\int^\infty_{\frac{\xl^2 \mpr^2}{1 - \xi}}
   \frac{\mathd Q^2}{Q^2}   \int_{z^{(1)}_{\min}}^{z^{(2)}_{\min}}
   \frac{\mathd z}{z} \,.
\end{equation}

In the case $M^2 (z) = \muF^2$ we do not have any restriction on $z$
from the first theta function, and the $z$ integrals are given by
\begin{equation} \theta (\muF^2 - Q^2) \int_{\xi}^1 \frac{\mathd z}{z} -
   \int_{z^{(2)}_{\min}}^1 \frac{\mathd z}{z} = \theta (\muF^2 - Q^2)
   \int_{\xi}^{z^{(2)}_{\min} } \frac{\mathd z}{z} - \theta (Q^2 - \muF^2)
   \int_{z^{(2)}_{\min}}^1 \frac{\mathd z}{z}\,.
\end{equation}
In this case, if the condition (\ref{eq:condq2z2}) holds, we have
  $z^{(2)}_{\rm min}=\xi$,
  and the first integral vanishes. Thus, in this case, the minimum value
  of $Q^2$ is
  \begin{equation}
    Q^2=\min\left(\frac{\mpr^2\xl^2}{1 - \xi}, \muF \right)\,,
  \end{equation}
  and the integration is written as
\begin{equation}
  \int_{\xl{}}^1 \frac{\mathd \xbj{}}{\xbj{}}
  \left[\int^{\muF^2}_{\min(\frac{\xl^2 \mpr^2}{1 - \xi},\muF^2)}
    \frac{\mathd Q^2}{Q^2}\int_{\xi}^{z^{(2)}_{\rm min}}\frac{\mathd z}{z}
    -\int^\infty_{\muF^2} \frac{\mathd Q^2}{Q^2}\int_{z^{(2)}_{\rm min}}^1 \frac{\mathd z}{z}
    \right]\,. \label{eq:altintmu2}
\end{equation}
The second term in Eq.~(\ref{eq:altintmu2}) is present if the $Q^2$
integration is interpreted as an oriented integral, and absent
otherwise.

The $z$ integration of Eq.~(\ref{eq:leptonpdf1}) written in terms of
the variables illustrated here was easily performed using MAXIMA
\cite{maxima}. We do not report here the lengthy result.  In order to
evaluate it with sufficient accuracy it must be implemented in
quadruple precision in the fortran code, in order to avoid sizable
rounding errors.

\section{The ${\cal O}(\aem^3)$ term}\label{app:aem3}
\begin{figure}[htb]
  \centering
  \includegraphics[width=0.32\linewidth]{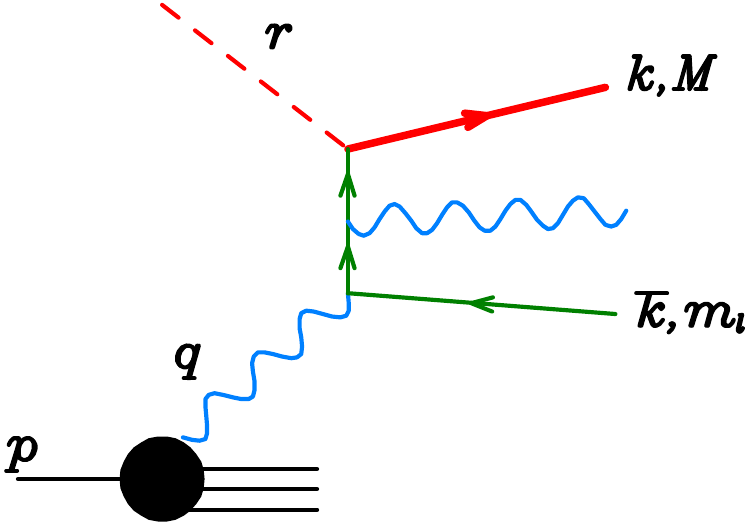}
  \caption{\label{fig:alpha3} A diagram contributing at order $\aem^3$
    with up to three powers of $\log(M^2/\Lambda^2)$ to the probe process.}
\end{figure}

Terms of order $\aem^3$ arise from graphs where one further photon
emission is allowed from the lepton, as the one illustrated in
Fig.~\ref{fig:alpha3}.
We need to compute the leading double-logarithmic term in the
leptonic tensor for this process. We have
\begin{equation} L^{\mu \nu} = L_1 \left( - g^{\mu \nu} + \frac{q^{\mu} q^{\nu}}{q^2}
   \right) + L_2 \frac{1}{r \cdot q} \left( r^{\mu} - \frac{q \cdot r
     q^{\mu}}{q^2} \right) \left( r^{\nu} - \frac{q \cdot r
     q^{\nu}}{q^2} \right) . \end{equation} The double logarithmic
region requires that $|q^2| \ll M^2$, and from analyticity
considerations we infer that $L_2$ must be proportional to $q^2$ for
small $q^2$ (up to logarithms), since $L^{\mu \nu}$ is the expectation
value of a product of currents, and thus cannot have poles in
$q^2$. Likewise, we must have that in
\begin{equation} L_1 \frac{q^{\mu} q^{\nu}}{q^2} + L_2 q \cdot r \frac{q^{\mu} q^{\nu}}{q^4}\,,
\end{equation}
the $q^2$ singularity must cancel. Thus, for small $q^2$
\begin{equation}
L_2\approx - L_1 \frac{q^2}{q \cdot r}.
\end{equation}
Therefore, in our limit
\begin{equation} \label{eq:Lmunusimp}
  L^{\mu \nu} = L_1 \left[ - g^{\mu \nu} + \frac{r^{\mu} q^{\nu} + r^{\nu}
      q^{\mu}}{r \cdot q} - r^{\mu} r^{\nu} \frac{q^2}{(q \cdot r)^2} \right],
\end{equation}
In order to evaluate $L_1$, we notice that
the cross section for an almost on-shell
photon to inclusively produce the heavy fermion is given in terms of the leptonic tensor
as
\begin{equation}
  \sigma = \frac{1}{2} \frac{1}{(-4 r \cdot q)}  (- L^{\mu}_{\mu}),
\end{equation}
where the first factor is for the spin average, the second is the flux
factor, and the third is the contraction of the leptonic tensor with the spin
projection. We now write $\sigma$ using the factorization formula twice
\begin{equation} \sigma = \frac{1}{(-8 r \cdot q)}  (- L^{\mu}_{\mu}) = \left(
   \frac{\alpha^{}}{2 \pi} \right)^2 \int_{Q^2}^{M^2} \frac{\mathd
   q_1^2}{q_1^2} \Plgamma (y_1) \mathd y_1 \int_{q_1^2}^{M^2} \frac{\mathd
   q_2^2}{q_2^2} P_{l l} (y_2) \mathd y_2 \sigma_B (r, q y_1 y_2),
\end{equation}
where $\sigma_B (r, q z y)$ is the Born cross section
\begin{equation} \sigma_B (r, q y_1 y_2) = \pi \delta ((r - q y_1 y_2)^2 - M^2) =
  \frac{\pi}{(-2 r \cdot q)} \delta (y_1 y_2 - \zl) ,
\end{equation}
and we have used $\zl\approx -M^2/(-2 r \cdot q)$ in the low $Q^2$ limit.
Thus
\begin{equation} - L^{\mu}_{\mu} = 4 \pi \left( \frac{\alpha^{}}{2 \pi} \right)^2 
   \frac{1}{2} \log^2 \frac{M^2}{Q^2}  \int^{} \Plgamma (y_1) \mathd y_1
   \int P_{l l} (y_2) \mathd y_2 \delta (y_1 y_2 - \zl) .
 \end{equation}
We define the convolution of splitting functions as
\begin{eqnarray}
  \Plgammatwo(\zl) & \equiv & \int \mathd y_1 \mathd y_2 \Plgamma (y_1) \Pll
  (y_2) \delta (y_1 y_2 - \zl) = \int_{\zl}^1 \frac{\mathd y_2}{y_2} \Plgamma \left( \frac{\zl}{y_2}
  \right) \Pll (y_2)\nonumber \\
  & = & \int_{\zl}^1  \frac{\mathd y_2}{y_2} \Plgamma \left( \frac{\zl}{y_2}
  \right) \Pll (y_2) - \int_0^1 \mathd y_2 \Plgamma (\zl) \Pll (y_2)\nonumber\,,
\end{eqnarray}
where 
\begin{equation}
  \Plgamma(y_1)=y_1^2+(1-y_1)^2,\quad\quad \Pll=\left( \frac{1+y_2^2}{1-y_2} \right)_+\,.
\end{equation}
Note that in the last line the plus-prescription on $\Pll$ is redundant since both expressions are finite.  
An explicit evaluation gives 
\begin{eqnarray}
  \Plgammatwo(\zl) 
  & = & (1 - 2 \zl + 4 \zl^2) \log \frac{1}{\zl} + \frac{1}{2}  (1 - \zl^2) (2 \zl^2 +
        2 \zl - 1) \nonumber \\
  && \phantom{aaaaaaa}+ \Plgamma (\zl) \frac{1}{2} (\zl^2 + 2 \zl + 4 \log (1 - \zl)) \,.
\end{eqnarray}
Finally we find
\begin{equation}
  L_1 = \frac{(-L^\mu_\mu)}{2}=\pi \left( \frac{\alpha^{}}{2 \pi} \right)^2 \log^2 \frac{M^2}{Q^2}
  P^{(2)}_{l \gamma} (\zl)\,,
\end{equation}
where we have used Eq.~(\ref{eq:Lmunusimp}).
From Eqs.~(\ref{eq:Lmunusimp}),~(\ref{eq:Wmunu}) and~(\ref{eq:FL}), dropping subleading terms 
we get
\begin{equation} W_{\mu \nu} L^{\mu \nu} = F_2 L_1 \frac{1}{x_{\tmop{bj}}}  \frac{1}{z^2} (1
   + (1 - z)^2) = F_2 L_1 \frac{1}{x_{\tmop{bj}}z}  \Pgammaq (z)\,.
 \end{equation}
 From Eq.~(\ref{eq:sigma3}) we get
\begin{eqnarray}
  \sigma & = & (4 \pi \alpha) \int \frac{\mathd E_{\tmop{cm}}^2}{2 \pi} 
  \frac{1}{4 p \cdot r}  \frac{1}{16 \pi^2 E_{\tmop{cm}}^2}  \int_x^{1 -
               \frac{2 x \mpr}{E_{\tmop{cm}}}} \mathd z \nonumber \\
  &\times& \int \frac{\mathd Q^2}{Q^2} 4 \pi
  F_2 \frac{1}{x_{\tmop{bj}}} \frac{1}{z^2} z \Pgammaq (z) \left[ \pi
  \left( \frac{\alpha}{2 \pi} \right)^2 \right] \log^2 \frac{M^2}{Q^2} P_{l
  \gamma}^{(2)} (\zl) \nonumber \\
  & = & \left( \frac{\alpha}{2 \pi} \right)^3 \frac{\pi}{2 S} \int
  \frac{\mathd x}{x} \frac{\zl }{x_l} \int_x^{1 - \frac{2 x
  \mpr}{E_{\tmop{cm}}}}  \frac{\mathd z}{z} \int_{\frac{\mpr^2 x^2}{1 -
  x}}^{\frac{E_{\tmop{cm}}^2 (1 - z)}{z}} \frac{\mathd Q^2}{Q^2} F_2 z
  \Pgammaq (z) \log^2 \frac{M^2}{Q^2} \Plgamma^{(2)} (\zl) \;. \phantom{aaa}
\end{eqnarray}
Since the parton model formula for the cross section reads
\begin{equation}
  \sigma = \int \mathd x \fl (x, M^2) \pi \delta (S x - M^2) = \frac{\pi}{S}
  \fl (x_l, M^2),
\end{equation}
the dominant $\aem^3$ contribution to the PDF, that
we denote $\fonel$, is given by
\begin{eqnarray}\label{eq:fonel}
  \xl\fonel (\xl, \muF^2) &=& \left( \frac{\alpha}{2 \pi} \right)^3 \int
                              \frac{\mathd x}{x} \zl \int_x^1 \frac{\mathd z}{z} \int_{\mpr^2}^{\muF^2}
                              \frac{\mathd Q^2}{Q^2} \nonumber \\
  && \phantom{aaaaaaaaaaaaaaaa} F_2(\xbj,Q^2) z
\Pgammaq (z) \frac{1}{2} \log^2 \frac{M^2}{Q^2} \Plgammatwo (\zl) \;,
\end{eqnarray}
where we have neglected the correction of
order $\mpr/\ecm$ in the upper limit of the $z$ integration, and performed
a simplification of the limits in the $Q^2$ integration allowed by our
target accuracy.

If we compare Eq.~(\ref{eq:fonel}) with Eq.~(\ref{eq:leptonpdf0}) we
immediately see that, if we consider only the leading logarithmic term
of Eq.~(\ref{eq:leptonpdf0}), the two equations are related by the
replacement
\begin{equation}
  \left( \frac{\alpha}{2 \pi} \right)^2 \frac{1}{2} \log^2 \frac{M^2}{Q^2} \Plgammatwo (\zl)
  \Leftrightarrow
  \left( \frac{\alpha}{2 \pi} \right) \log \frac{M^2}{Q^2} \Plgamma (\zl)\,.
\end{equation}
In fact, with the same procedure used in this section we could have
computed the leading logarithmic term in
formula~(\ref{eq:leptonpdf0}), the only difference being that we would
have had a single logarithmic integral, and thus a single log instead
of half a log squared, and a single splitting function to find a
lepton in the photon, instead of the convolution of two splitting
functions that we computed here.

Eq.~(\ref{eq:fonel}) can also be written, performing a change of variables, and using the
parton model formula for $F_2$, in the form
\begin{eqnarray}\label{eq:fonel1}
  \fonel (\xl, \mu_F^2) &=&
  \left( \frac{\alpha}{2 \pi} \right)^3 \int_0^1 \mathd \xbj \int_0^1 \mathd
   z P_{\gamma q} (z) \int_0^1 \mathd \zl P_{l \gamma}^{(2)} (\zl) \delta
                            (\xbj z \zl - \xl) \nonumber \\
 && \phantom{aaaaaaaaaaa}\int_{\mpr^2}^{\muF^2} \frac{\mathd Q^2}{Q^2}
   \sum_if_i(\xbj, Q^2) c^2_i  \frac{1}{2} \log^2 \frac{M^2}{Q^2} .
\end{eqnarray}
This form is very suggestive, since it is the convolution of three
leading order splitting functions combined with the (ordered)
integration of three intermediate scales.  In fact, this equation can
be obtained by iterating the first three rungs of the integral form of
the leading-order Altarelli-Parisi equation, under the assumption that
$\fonel $ vanishes when $\muF\approx\mpr$. This is certainly the case,
since $\fonel $ is of order $\aem^3$ without logarithmic enhancement
in this limit.

\begin{figure}[htb]
  \centering
  \includegraphics[width=\linewidth]{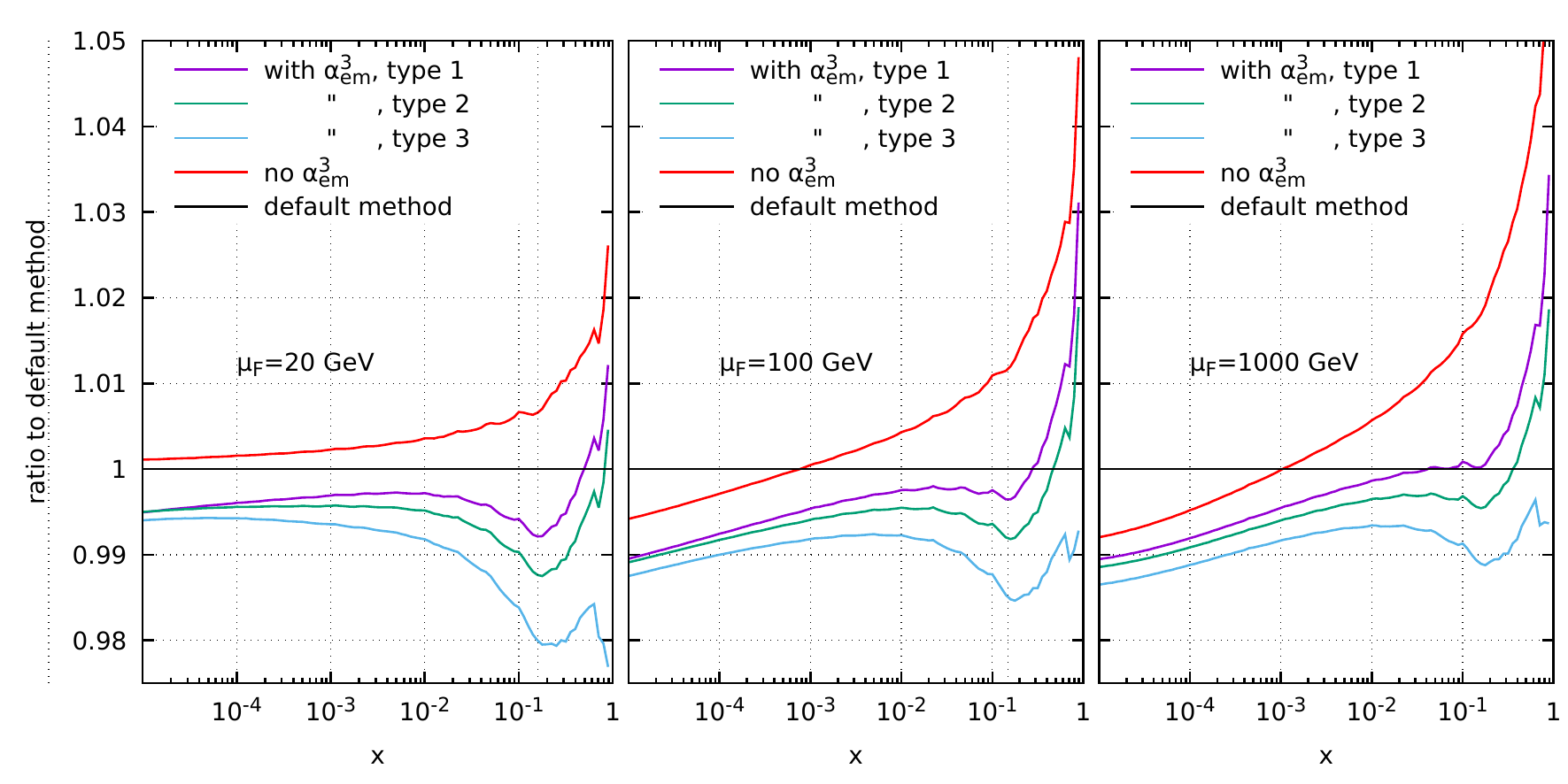}
  \caption{\label{fig:aem3} Comparison of the different methods to include the effects of order $\aem^3$, see text for more details.}
\end{figure}

The calculation described in this appendix could be used to compute
directly the leptons PDFs at any (perturbative) scale, without making
use of step~\ref{enum:backevol} of section \ref{sec:pdfset}. It is
therefore interesting to compare what we obtain with this method with
respect to our default one. The comparison is shown in
Fig.~\ref{fig:aem3}.
The lines shown in the plots are the result of our main formula for
the lepton densities, Eq.~(\ref{eq:leptonpdf1}), with (black) or
without (red) the terms of order $\aem^3$ given in this appendix. We
show results at three different values of the factorization scale, and
normalize them to our default procedure, described in
item~\ref{enum:backevol} in Sec.~\ref{sec:pdfset}, corresponding to
the central value of the \LUXlep{} set. We have implemented the $\aem^3$
contribution using three different variations (denoted as ``types'' in
the figure), that should all agree up to higher-order corrections. In
all the three types we have used the integration limits of
formula~(\ref{eq:sigma3}). In type 1 we use formula~(\ref{eq:sigma3})
as is, in type 2 we replace one of the two powers of logarithms in
Eq.~(\ref{eq:sigma3}) with
\begin{equation}
  \log \frac{M^2}{Q^2}\rightarrow \log\frac{\muF^2}{ Q^2(1-\zl)\zl+\ml^2},
\end{equation}
and for type 3 we make the same replacement for both powers of the
logarithm. This form of the logarithm is the one that we have in
formula~(\ref{eq:leptonpdf1}). Of course, without making a more
detailed calculation we cannot know the true form of the argument in
the logarithm. We use this artifact just to gauge the sensitivity to
subleading effects.  First of all, we notice that the $\aem^3$
correction, irrespective of the method used to compute it, does not
seem to be as important as the NLO correction included in
formula~(\ref{eq:leptonpdf1}), that is suppressed by the absence of a
large logarithm (which in our counting is equivalent to a power of
$\as$). While the latter is of order 10\%{}, the correction that we
computed here is not much larger than 1\%{}, if one excludes the very
large $x$ region.

All the four methods that we have considered to include $\aem^3$
effects (i.e. type 1, 2 and 3 plus our default method) seem to reduce
the density evaluated with formula~(\ref{eq:leptonpdf1}) at large
$x$. This is consistent with the expectation that the electromagnetic
radiation from leptons should be particularly important at large
values of $x$, where it tends to soften the lepton distributions. In
all cases, our default method seems to be more effective in doing so
than the calculation performed in this appendix. This is perhaps due
to the fact that mixed QED-QCD splitting kernels are included with our
default method, but not in types 1-3.  We also notice that subleading
terms, estimated as the difference between types 1, 2 and 3, are not
negligible if compared to the full magnitude of the effect (which is
the difference with respect to the result without $\aem^3$ effects)
being several tens of percent of the overall $\aem^3$ effect. The same
can be said about the difference of the default result with respect
type 1, 2 and 3. There we observe a deviation that is also not
negligible in the small $x$ region.

The variation between the default method and type 1, 2 and 3 could be
added to our calculation as a further source of uncertainty arising
from higher-order effects.  Since the uncertainty that we find here is
considerably smaller than the one obtained with the method described
as item (HO) of Section~\ref{sec:pdfset} we do not include it, also
reassured by the fact that even if we did include it, the final error
estimate would not change substantially.

\bibliographystyle{JHEP}
\bibliography{LeptonPDF}

\end{document}